\documentclass[12pt, draftclsnofoot, onecolumn]{IEEEtran}
\usepackage{graphicx, times, amsmath, amsfonts, comment,amsthm}

\usepackage{ enumitem}
\usepackage{amssymb, epstopdf, soul, color}
\usepackage[noend]{algorithmic}
\usepackage{algorithm, array}

\newcommand{\beq}{\begin{equation}}
\newcommand{\eeq}{\end{equation}}
\newcommand{\tbf}{\textbf}
\newcommand{\tit}{\textit}
\newcommand{\mbf}{\boldsymbol}

\newcommand{\ud}{\mathrm{d}}
\newcommand{\sgn}{\mathrm{sgn}}

\theoremstyle{plain}
\newtheorem{lemmacounter}{Theorem}
\newtheorem{proposition}[lemmacounter]{Proposition}
\theoremstyle{plain}
\newtheorem{corocounter}{Theorem}
\newtheorem{corollary}[corocounter]{Corollary}
\theoremstyle{plain}

\newcommand {\Ebb}{\mathbb{E}}

\newcommand{\Lbb}{\mathbb{L}}
\newcommand {\Rbb}{\mathbb{R}}

\newcommand {\Bcal}{\mathcal{B}}

\newcommand {\Ical}{\mathcal{I}}
\newcommand {\Kcal}{\mathcal{K}}
\newcommand {\Lcal}{\mathcal{L}}
\newcommand {\Mcal}{\mathcal{M}}

\begin{document}
\title{Analysis of SINR Outage in Large-Scale Cellular Networks Using Campbell's Theorem and Cumulant Generating Functions
}

\author{
\IEEEauthorblockN{Sudarshan Guruacharya, Hina Tabassum, and Ekram Hossain\thanks{The authors are with the Department of Electrical and Computer Engineering at the University of Manitoba, Canada.}}

}\maketitle


\begin{abstract}
The signal-to-noise-plus-interference ratio (SINR) outage probability is one of the key performance parameters of a wireless cellular network, and its analytical as well as numerical evaluation has occupied many researchers. Recently, the introduction of stochastic geometric modeling of cellular networks has brought the outage problem to the forefront again. A popular and powerful approach is to exploit the available moment generating function (or Laplace transform) of received signal and interference, whenever it exists, by applying the Gil-Pelaez inversion formula. However, with the stochastic geometric modeling, the moment generating function may either be too complicated to exist in closed-form or at worst may not exist. Toward this end, in this paper, we study two alternate ways of evaluating the SINR outage. In the first case, we  emphasize the significance of calculating cumulants over  moments and exploit the fact that the cumulants of point processes are easily calculable using Campbell's theorem. The SINR outage is then analytically characterized by  Charlier expansion based on Gaussian and Student's $t$-distributions and their associated Hermite and Krishnamoorthy polynomials. In the second case, we  exploit the saddle point method, which gives a semi-analytical method of calculating the SINR outage, whenever the cumulant generating function of received signal and interference exists. For the purpose of demonstration, we apply these techniques on a downlink cellular network model where a typical user experiences a coordinated multi-point transmission, and the base stations are modeled by homogeneous Poisson point process. For the convenience of readers, we also provide a  brief overview of moments, cumulants, their generating functions, and Campbell's theorem, without invoking measure theory.  Numerical results illustrate the accuracy of the proposed mathematical approaches.
\end{abstract}




\section{Introduction}
Signal-to-noise-plus-interference ratio (SINR) outage probability (i.e., the probability of the SINR falling below a predefined threshold)  is one of the primary performance metrics for the analysis of a wireless communication system. The simplicity of its definition as well as its connection with other performance parameters, such as bit/symbol error rate and ergodic capacity, make it of significant interest to researchers. Till date, there have been numerous researches that have focused on analyzing the exact or approximate outage probability of wireless systems in diverse network settings and under varying modeling assumptions such as multi-path channel fading \cite[references therein]{Annamalai2001}, aggregate interference, and distance geometry of transmitters/receivers.  Based on the sources of randomness, we  can categorize various modeling scenarios as follows:
\begin{enumerate}[label=(\Alph*)]
\item \tit{Uncertainty due to multi-path channel fading and number of interferers}: In this case, both the desired channel and the interfering channel are assumed to undergo multi-path fading. The number of interferers is also random. Only the distance geometry between the receivers and the transmitters is considered to be deterministic.
\item \tit{Uncertainty due to aggregate interference and distance-based path-loss}: In this case, multi-path channel fading is assumed to be absent. The uncertainty is due to the random location of the transmitters and the receivers, and the unknown number of interferers.
\item \tit{Uncertainty due to multi-path channel fading, aggregate interference, and distance-based path-loss}: This case incorporates the uncertainty due to multi-path channel fading, aggregate interference, and the distance geometry of the transmitters and the receivers. 
\end{enumerate}
Evidently, the outage analysis of the latter models become  more complicated due to the combined effect of different sources of uncertainty. Also, the resulting interference distributions of the latter cases tend to be heavy tailed and thus many of the well-known approximations may not provide reliable and accurate performance characterization of the system.

\subsection{Background Work}
A very popular and powerful approach to calculate the outage is to utilize the moment generating function (MGF) of the relevant random variables. 
Provided the MGF of the received signal power and aggregate interference, \tit{Gil-Pelaez's inversion formula} is typically used to find the SINR outage at a receiver~\cite{Gil-Pelaez1951}. This involves calculating an integral involving the MGF, provided that the MGFs of related random variables exist\footnote{Recently, an MGF-based capacity evaluation technique has also been given in \cite{Hamdi2010}.}. Recently, the introduction of stochastic geometry in the modeling of wireless cellular communication systems has brought the conventional outage (or coverage) problem to the forefront again \cite[references therein]{Haenggi2012}. The stochastic geometry involves modeling the transmitters and receivers as a point process in two dimensional (2-D) plane. Due to its analytical tractability, the most frequently used point process is the Poisson point process (PPP)
\cite[references therein]{Andrews2011,Dhillon2012, ElSawy2013}. However, even for  PPP, where a typical receiver  associates with the nearest base station (BS), 
\begin{itemize}
\item the analytically tractable SINR outage formula relies on the so called ``Rayleigh fading trick'' \cite{ElSawy2013}. Unfortunately, this trick relies on the fading of the useful link  to be Rayleigh and cannot be extended to other fading scenarios. 
\item the interference distributions are often intractable and tend to have high skewness and kurtosis (heavy tailed), due to the fact that the transmitters can be arbitrarily close to a receiver\footnote{A simple remedy is to introduce an exclusion region around a receiver, the size of which will influence the heaviness of the interference distribution's tail. A more sophisticated remedy is to introduce repulsive point processes that allows more realistic modeling, but at the expense of analytical tractability.}. Consequently, exact/approximate closed-form SINR outage (or coverage) expressions can rarely be derived. 
\end{itemize}

The integrals required to evaluate the outage can be too complicated to be solved in closed-form. As such, either these integrals are expressed in terms of special functions (whose stability may  be unknown)  or  are evaluated using numerical integration (by which qualitative understanding is lost). It is therefore worthwhile to examine how outage can be approximated in an efficient manner without explicit integrations. One such technique, which was investigated in the late 20th century, is the saddle point approximation (SPA) method. It involves evaluating the cumulant generating function (CGF) at a single point, called the saddle point, where most of the value of integral is concentrated. The saddle point method has previously been utilized  to evaluate detection probability \cite{Helstrom1978}, error probability \cite{Yue1979}, and outage probability \cite{Guruacharya2016}. The work in \cite{Guruacharya2016} exploited the \tit{Lugannani-Rice formula} to compute the SINR outage probability for well-known fading distributions. 

Unfortunately, the MGF of a random variable may not always exist. This is the case for log-normal and most composite fast and slow-fading distributions. Furthermore, for situations where {\bf MGF/CGF does not exist, it may be legitimate to ask if one can directly compute the SINR outage from the moments that usually exist even when the MGF may not.} This is an extremely general problem and has been well studied in Mathematics. The problem of finding the distribution given the moments of a random variable is known as the \tit{problem of moments} \cite{Akhiezer1965}. In this context, rational functions (also known as \tit{Pade} approximants) have been used as an analytical approximation of the MGF when the moments of a random variable are given, but whose MGF may not exist \cite{Amindavar1994a,Amindavar1994b,Stokes1998}. Specifically, the inverse Laplace transform of the partial fraction decomposition of the rational function gives the outage probability as a sum of weighted exponentials. Unfortunately, this method is numerically unstable when moments of very high orders are considered. It is also difficult to intuit how the moments directly affect the outage. Lastly, this method is unreliable for heavy-tailed distributions.

\subsection{Paper Contributions}
To this end, the contributions of this  paper are listed herein.
\begin{enumerate}
\item We review and exploit new mathematical techniques for the evaluation of SINR outage probability of a receiver in a large-scale wireless cellular network. The techniques are general enough to be applied to stochastic geometry-based network models. As such, to demonstrate the application of the presented mathematical techniques, we  choose a stochastic geometry-based cellular network model, where BSs are distributed according to a homogeneous Poisson Point Process (PPP), and where a typical user experiences a downlink coordinated multi-point (CoMP) transmission. The outage is derived considering all three cases (i.e., Cases (A), (B), and (C) listed in the beginning of this section).

\item We propose an efficient approximation for SINR outage probability using saddle point methods.  In this approach, we need to evaluate a single point of the function to be integrated, called the saddle point, based on which outage can be obtained. The technique relies on the CGF representation of a random variable. We propose a general version of the saddle point method due to Wood, Booth and Butler, from which the \tit{Lugannani-Rice formula} can be derived as a special case. The general approach allows us to tackle distributions with heavy tails. For demonstration purposes, we  show the application of saddle point approximation in various network scenarios.

\item For situations where  MGF or CGF is not available, we propose the use of orthogonal polynomials expansion, which relies on the moments of the received signal and interference random variables, to derive the SINR outage probability of a general wireless communication system. Such expansions are familiar to us as generalized Fourier series, and in probability are known as \tit{Charlier} expansion. The special case with Hermite polynomials is known as \tit{Gram-Charlier Type A} expansion. However, the method is much more general, and we give an exposition on how this method can be utilized for the calculation of SINR outage probability. We provide an analytical expression to compute the signal-to-interference ratio (SIR) outage probability directly from the moments using Hermite polynomials and Krishnamoorthy polynomials.

\item Although we exploit orthogonal polynomial expansion that rely on moments, we establish and emphasize the significance of calculating the cumulants prior to moments and then applying the orthogonal polynomial expansion. In this regard, we show that the cumulants of the point processes are easily calculable via \tit{Campbell's Theorem}, and are often in an analytically tractable form. This fact, however, does not seem to be highlighted/utilized in any of the pioneering stochastic geometric frameworks such as \cite{Haenggi2012}. 
\end{enumerate}

The rest of the paper is structured as follows: Section II discusses the preliminary mathematical ideas necessary to understand the paper. Section III discusses the application of Gil-Pelaez formula, highlights the significance of evaluating cumulants prior to moments, and outlines the considered network model we will be dealing with throughout the paper. Section IV gives a brief overview of the moment problem, the reconstruction of a distribution given the cumulants (and in turn moments), and the corresponding SIR outage calculation. Section V  describes the saddle point method. Section VI considers the application of SPA in different wireless modeling scenarios. Numerical results are given in Section VII while Section VIII concludes the paper.


\section{Mathematical Preliminaries}
In this section, we briefly review some mathematical concepts related to moments, cumulants, and their transformations, that serve as a prerequisite for readers. A more extensive and in-depth discussion of moments, cumulants and their generating functions can be found in \cite[Ch. 3]{StuartOrd1994}. Materials on moment-cumulant transformations and Bell polynomials can be found in \cite[Ch. 3.3]{Comtet1970}, while the details of Campbell's formula can be found in \cite{Rice1944, Lowen1990}.

\subsection{MGF, CGF, and Their Properties }
Let the MGF, or in general the Laplace transform, of a random variable $X$ be given by $\Mcal_X(t) = \Ebb[e^{-t x}]$, then its CGF is given by the logarithm of the MGF $\Kcal_X(t) = \log \Mcal_X(t)$.  $\Mcal_X(t)$ exists if it is finite around the neighborhood of zero, i.e., if there exists an $h > 0$ such that $\forall t \in (-h,h)$, we have $\Mcal_X(t) < \infty$. If $\Mcal_X(t)$ exists, the largest open interval $U$ around zero such that $\Mcal_X(t) < \infty$ for $t\in U$ is referred to as the convergence strip of the MGF of $X$. Also, if $\Mcal_X(t)$ exists, then all positive moments of $X$ exist and CGF exists as well. 

Some of the fundamental properties of MGF and CGF include:~(i) if  $Y = aX + b$, where $a$ and $b$ are some constants, then we can express the MGF and CGF of $Y$ as $\Mcal_Y(t) = e^{-bt} \Mcal_X(at)$ and $\Kcal_Y(t) = \Kcal_X(at) - bt$, respectively, (ii)~if $Z = X + Y$ such that $X$ and $Y$ are independent, then $\Mcal_Z(t) = \Mcal_X(t) \Mcal_Y(t)$ and $\Kcal_Z(t) = \Kcal_X(t) + \Kcal_Y(t)$, and
(iii)~if $Y = \sum_{i=1}^N X_i$ is a compound distribution, where $X_i$ are independent and identically distributed (IID) and $N$ is a discrete random variable, then $\Mcal_Y(t) = \Mcal_N(-\log \Mcal_X(t))$ and $\Kcal_Y(t) = \Kcal_N(-\Kcal_X(t))$. This can be shown by first conditioning on $N$, $\Mcal_{Y|N}(t) = \prod_{i=1}^N \Mcal_X(t) = \exp (N \log \Mcal_X(t))$. Now, taking expectation with respect to $N$, we get $\Mcal_Y(t) = \Ebb_N[e^{N \log \Mcal_X(t)}] = \Mcal_N(-\log \Mcal_X(t))$.

\subsection{Moments, Cumulants, and Their Properties}
The $n$-th moment of $X$ is given by 
$\mu_n(X) = \int x^n f(x) \ud x.$
Clearly, the $n$-th moment exists if the integral is finite. If the MGF of $X$ exists, the moments can also be defined as
$ \mu_n (X) = (-1)^n  \lim_{t\rightarrow0} \frac{\ud^n}{\ud t^n} \Mcal_X(t). $
In other words, the $n$-th moment is the $n$-th coefficient in the Taylor expansion of $\Mcal_X(t)$, and $\Mcal_X(t)$ can be represented as
$\Mcal_X(t) = \sum_{n=0}^\infty  \mu_n \frac{(-t)^n}{n!}. $
Even if the moments of a random variable exists, its MGF may not always exist. Some of the important properties of moments are~(i) \tit{\bf Homogeneity}: If $c$ is some constant, then $\mu_n(cX) = c^n \mu_n(X)$, and (ii)~\tit{\bf Independence}: If $X$ and $Y$ are independent,  $\mu_n(XY) = \mu_n(X) \mu_n(Y)$.

Similarly, the $n$-th cumulant of $X$ is given by 
$ \kappa_n(X) = (-1)^n \lim_{t\rightarrow0} \frac{\ud^n}{\ud t^n} \Kcal_X(t).$
Similar to the moments, the $n$-th cumulant is the $n$-th coefficient in the Taylor expansion of $\Kcal_X(t)$ and $\Kcal_X(t)$ can be represented as
$ \Kcal_X(t) = \sum_{n=0}^\infty  \kappa_n \frac{(-t)^n}{n!}. $
Some of the important properties of cumulants are 
(i)~\tit{\bf Additivity}: If $X$ and $Y$ are independent, then $ \kappa_n(X+Y) = \kappa_n(X) + \kappa_n(Y)$,
(ii)~\tit{\bf Invariance}: If $c$ is some constant, then $\kappa_1(X+c) = \kappa_1(X) + c$ and $\kappa_n(X+c) = \kappa_n(X)$ for $n \geq 2$,
(iii)~\tit{\bf Homogeneity}: If $c$ is some constant, then $\kappa_n(cX) = c^n \kappa_n(X). $

{\bf Remark:} When dealing with the sum of independent random variables, the additive property of the cumulants  makes them easier to work with compared to moments. As such, cumulants naturally occur in the study of central limit theorems.

 \subsection{Transformation of Moments and Cumulants}
The moments and cumulants can be transformed into each other as detailed below:
 \begin{align*} 
 \mu_n &= \sum_{k=1}^n B_{n,k}(\kappa_1, \ldots, \kappa_{n-k+1}), \\
 \kappa_n &= \sum_{k=1}^n (-1)^{k-1} (k-1)! B_n(\mu_1, \ldots, \mu_{n-k+1}).
 \end{align*}
 Here the $B_{n,k}(x_1,x_2,\dots,x_{n-k+1})$ are the partial exponential Bell polynomials defined as
 \[ B_{n,k}(x_1,x_2,\dots,x_{n-k+1}) = \sum{n! \over j_1!j_2!\cdots j_{n-k+1}!}
\left({x_1\over 1!}\right)^{j_1}\left({x_2\over 2!}\right)^{j_2}\cdots\left({x_{n-k+1} \over (n-k+1)!}\right)^{j_{n-k+1}}, \]
where the sum is taken over all sequences $j_1, j_2, \ldots, j_{n-k+1}$ of non-negative integers such that these two conditions are satisfied: $ j_1 + j_2 + \cdots + j_{n-k+1} = k$ and $j_1 + 2 j_2 + \cdots + (n-k+1)j_{n-k+1} = n$.
Equivalently, the partial exponential Bell polynomials can also be defined recursively as 
\[ B_{n,k}(x_1,\dots,x_{n-k+1}) = \sum_{i=1}^{n-k+1} \binom{n-1}{i-1} x_{i} B_{n-i,k-1}(x_1,\dots,x_{(n-i)-(k-1)+1}), \]
where $B_{0,0} = 1$, $B_{n,0} = 0$ for  $n \geq 1$ and $B_{0,k} = 0$ for $k \geq 1$. This recursive definition is useful for computational purposes. The Bell polynomials are used in the study of set partitions, and occur often in applications, such as in the Faa di Bruno's formula.

\subsection{Campbell's Theorem for Poisson Point Process (PPP)}
Given a homogeneous PPP $\Phi$ of uniform intensity $\lambda$ defined over a scalar parameter {$x \in \Ical$ where $\Ical \subset \Rbb$}, let an aggregate process be given by
$ Y(x) = \sum_{x_i \in \Phi} h(x - x_i \,; \xi_i) $ 
 where $h(x; \xi)$ is a deterministic function conditioned on $\xi$ and $\xi_i$ are IID random variables. The variable $\xi$ is commonly used to model the random amplitude of the function $h(x)$ as $h(x ; \xi) =\xi h(x)$. {The number of points $N$ in the interval $\Ical$ is Poisson distributed, hence the CGF of $N$ is $\lambda |\Ical| (e^{-t} - 1)$. Likewise, the point $x$ is uniform randomly selected from the interval $\Ical$. Using the CGF property for compound distributions and de-conditioning on $x$, the CGF of $Y$ for such PPP is: }
  \[ \Kcal_Y(t) = \int_\Ical \Kcal_{N}(-K_{H}(t)) \frac{\ud x}{|\Ical|} =  \lambda \int_\Ical (\Ebb_H[e^{-t h(x; \, \xi)}] - 1) \ud x, \]
  where we use the fact that $e^{\Kcal_{H}(t)} = \Mcal_{H}(t) = \Ebb_H[e^{-t h(x;\, \xi)}]$. In order to find the cumulants, we first find the $n$-th derivative of $\Kcal_Y(t)$ with respect to $t$ as 
  \[ \frac{\ud^n}{\ud t^n} \Kcal_Y(t) = (-1)^n \lambda \int_\Ical \Ebb_H[ h(x; \xi)^n e^{-t h(x; \, \xi)}] \ud x. \]
Taking the $n$-th derivative of $\Kcal_Y$ as $t\rightarrow0$, we obtain the \tit{Campbell's formula}
$\kappa_n(Y)  = \lambda  \Ebb_H \int_\Ical [h(x; \xi)]^n \ud x. $
Given $h(x; \xi) =\xi h(x)$, the formula simplifies to a more useful form:
 \beq 
 \kappa_n(Y)  = \lambda \mu_n(\xi) \int_\Ical [h(x)]^n \ud x, 
 \label{eqn:campbell-formula}
 \eeq
 where $\mu_n(\xi)$ is the $n$-th moment of $\xi$. The Campbell's formula allows us to calculate the  cumulants of the aggregate directly without the need to compute its CGF and its derivatives. Given the cumulants of $Y$, we can find its moments using the transformation described in the previous subsection. The Campbell's formula can also be extended to non-uniform PPP as well as non-Poisson point process. However, for our current work, we only focus on uniform PPP.


\section{Evaluation of SINR Outage  and System Model}
In this section, we will first describe the general outage formulation and 
evaluation by applying {\em Gil-Pelaez inversion formula} using both the MGF and CGF approaches, which is valid for both uplink and downlink scenarios. The significance of deriving cumulants over moments is then highlighted. Finally, we describe the large-scale cellular network model of our interest, which is considered throughout the paper.

\subsection{General SINR Outage Evaluation}
Let us define the SINR of a wireless system as $\mathrm{SINR} = \frac{\sum_{i=1}^M X_i}{1 + \sum_{j=1}^N Y_j}$, where $X_i$ denotes the useful signal power, $Y_i$ denotes the power of the interference signal, and the noise power is normalized to unity. Every $X_i$ and $Y_i$ are assumed to be independent random variables. If we neglect the noise term, then the signal-to-interference-ratio (SIR) is given by 
$\mathrm{SIR} = \frac{\sum_{i=1}^M X_i}{\sum_{j=1}^N Y_j}.$ Consider $X = \sum_{i=1}^M X_i$, $Y = \sum_{j=1}^N Y_j$, and a given SIR threshold $\theta$, the SIR outage occurs when  $\theta Y> X$. Following Zhang's approach \cite{Zhang1995}, let us define a new random variable $\Omega = \theta Y - X$, then the SIR outage probability can  be given as: 
\beq 
P_{\mathrm{out}} = \mathrm{Pr}(\Omega > 0) = Q_\Omega(0), 
\label{eqn:outage-def}
\eeq
where $Q_\Omega$ is the complementary cumulative distribution function (CCDF) of $\Omega$. If we do not neglect the noise, then the SINR outage will be given by $P_{\mathrm{out}} = \mathrm{Pr}(\Omega > -\theta)$.
Given that $\Omega$ is a linear combination of independent random variables, we can  obtain the MGF of $\Omega$. The SINR outage probability in (\ref{eqn:outage-def}) can then be evaluated using Gil-Pelaez inversion formula as~\cite{Gil-Pelaez1951}:
\beq
Q_{\Omega}(\omega) = \frac{1}{2} - \frac{1}{\pi} \int_0^\infty \mathrm{Im}\{\Mcal_\Omega(\jmath t) e^{-\jmath t\omega}\} \frac{\ud t}{t},
\label{eqn:gil-pelaez-outage}
\eeq
where $\Mcal_\Omega(t)$ is the MGF of $\Omega$, $\mathrm{Im\{z\}}$ is the imaginary component of complex variable $z$, and $\jmath = \sqrt{-1}$.  The MGF  of $\Omega$ is given by $\Mcal_\Omega(t) = \Mcal_Y(\theta t)\Mcal_X(-t)$. 
If the signals $X_i$ and $Y_i$ are mutually independent, we further have $\Mcal_\Omega(t) = \prod_{i=1}^N \Mcal_{Y_i} (\theta t) \prod_{i=1}^M \Mcal_{X_i}(-t)$.

Given that $\Mcal_\Omega(\jmath t) = \exp \log{\Mcal_\Omega(\jmath t)} = \exp \Kcal_\Omega(\jmath t)$,  (\ref{eqn:gil-pelaez-outage}) can be restated in terms of CGF as $Q_{\Omega}(\omega) = \frac{1}{2} - \frac{1}{\pi} \int_0^\infty \mathrm{Im}\{ e^{\Kcal_\Omega(\jmath t) - \jmath t\omega}\} \frac{\ud t}{t},$  where 
\beq
\Kcal_\Omega(t) = \Kcal_Y(\theta t) + \Kcal_X(-t),
\label{eqn:CGF-omega-generic-case}
\eeq 
and if the signals $X_i$ and $Y_i$ are mutually independent, we further have $\Kcal_\Omega(t) = \sum_{i=1}^N \Kcal_{Y_i} (\theta t) + \sum_{i=1}^M \Kcal_{X_i}(-t).$

Note that, the $n$-th moment and cumulant of $\Omega$ can be found by $n$-fold differentiation of the MGF and CGF of $\Omega$, respectively. However, the multiplicative representation of the MGF of $\Omega$ in terms of MGFs of $X_i$ and $Y_i$ necessitates the application of Leibnitz's product rule for $M+N$ terms. The resulting formula for the $n$-th moment ends up being quite complicated due to multinomial series. On the other hand, the additive nature of $\Omega$ makes the calculation of cumulants of $\Omega$ much easier compared to its moments, as described in the following proposition.
\begin{proposition}[$n^{\mathrm{th}}$ Cumulant of $\Omega$] 
The $n$-th cumulant of $\Omega$ is given by
\beq 
\kappa_n(\Omega) = \theta^n \kappa_n(Y) + (-1)^n \kappa_n(X). 
\label{eqn:nth-cumulant-gamma-generic-model}
\eeq
\label{prop:cumulant-generic-model}
\end{proposition}
\begin{IEEEproof}
 Using the additivity and homogeneity properties of cumulants.
\end{IEEEproof}

\subsection{Representative Large-Scale Cellular Network Model}

\subsubsection{Spatial model} 
\begin{figure}[h]
\begin{center}
	\includegraphics[width=4in]{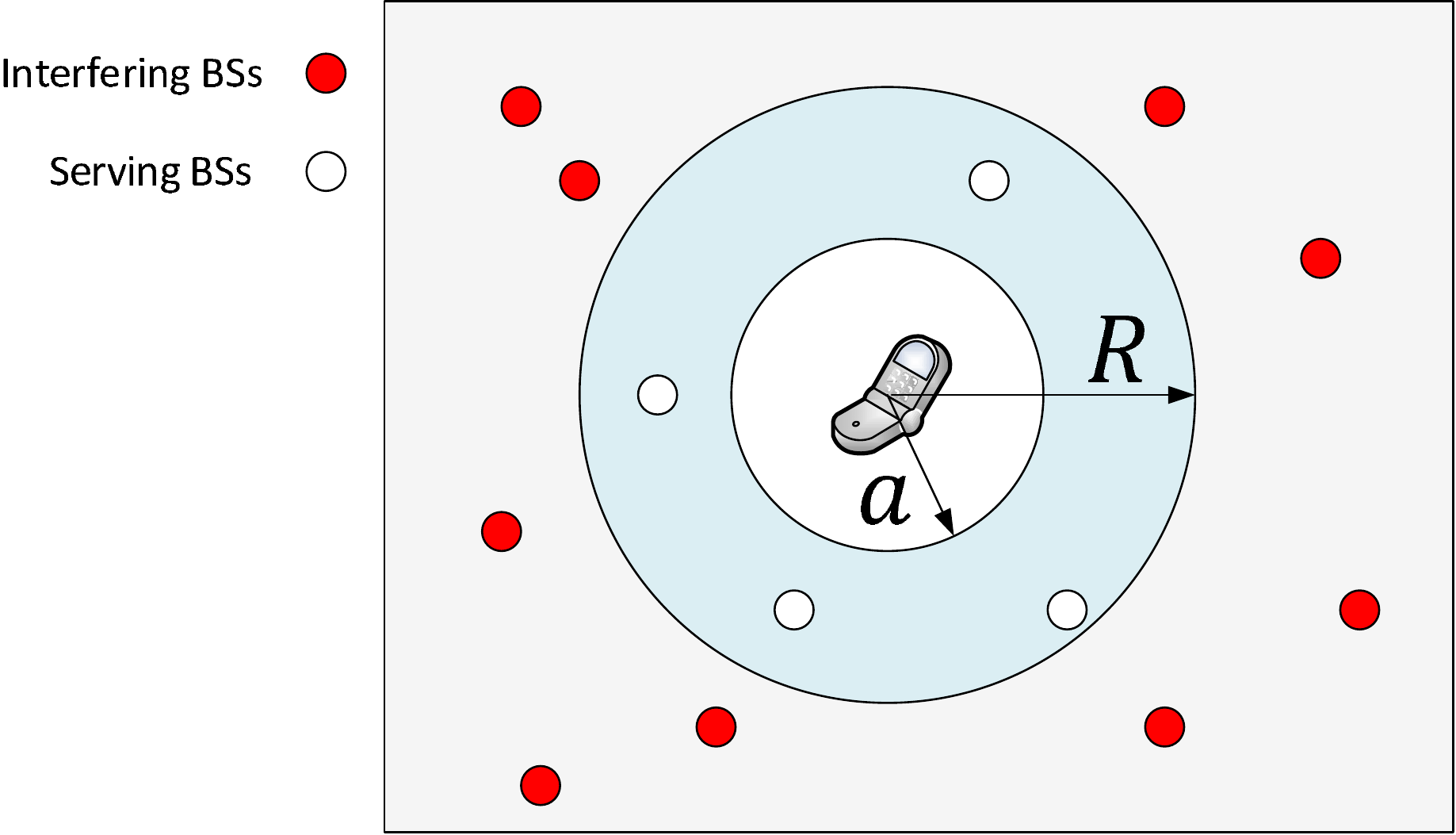}
	\caption{The exclusion region, the region of cooperation, and the region of interference. }
	\label{fig:different-zones}
 \end{center}
\end{figure}

Consider a single-antenna user equipment (UE) located at the origin, as shown in Fig. \ref{fig:different-zones}. Let single antenna BSs be scattered in 2-D plane according to homogeneous PPP of intensity $\lambda$. Consider an annular region $\Bcal^c$ centred at origin and with \tit{fixed} outer radius $R$ and inner radius $a > 0$, such that $\Bcal^c = \{r | a \leq r < R\}$. It is assumed that there are no BSs located within radius $r < a$, thus forming an \tit{exclusion region}. All BSs within $\Bcal^c$ are assumed to cooperate with each other to form coordinated multi-point (CoMP) transmission to a user equipment (UE) located at the origin, thus forming a \tit{cooperation region}. All BSs beyond $R$,  $\Bcal^{nc} = \{r| r \geq R\}$, act as interferers; thus $\Bcal^{nc}$ forms the \tit{interference region}. Since $\Bcal^{n}\cap \Bcal^{nc} = \varnothing$, the BSs in $\Bcal^{c}$ and $\Bcal^{nc}$ will both be PPP of intensity $\lambda$, as per the property of PPP\footnote{The same scenario can also be considered for the uplink, where the UE transmits a message, which is cooperatively detected by the BSs.}. 

\subsubsection{Received signal and interference model} The instantaneous received signal of the UE at the origin is modeled as: 
 \[v = \sum_{i\in\Phi_{\Bcal^c}} \sqrt{g_i} r_i^{-\frac{\alpha}{2}} s + \sum_{j\in\Phi_{\Bcal^{nc}}} \sqrt{g_j} r_j^{-\frac{\alpha}{2}} s_j + z, \] 
 where $s$ is the common message signal transmitted by all BSs in $\Bcal^{c}$ and $s_j$ are interfering signals from BSs in $\Bcal^{nc}$. The $r_i$ is the distance between $i$-th BS and the UE at the origin, the $g_i$ are IID channel gains, and $\alpha$ is the path-loss exponent {such that $\alpha > 2$}. Lastly, $z$ is the additive white noise. Let the variances $\mathrm{Var}(s) = \mathrm{Var}(s_i) = P$ and $\mathrm{Var}(z) = \sigma^2$. Assuming maximum-ratio-combining (MRC) at the UE, we can write the SINR as 
 \[ \mathrm{SINR} = \frac{\sum_{i\in\Phi_{\Bcal^{c}}} g_i P r_i^{-\alpha}}{\sum_{j\in\Phi_{\Bcal^{nc}}} g_j P r_j^{-\alpha} + \sigma^2}. \]
Assuming negligible noise, $X = \sum_{i\in\Phi_{\Bcal^{c}}} g_i P r_i^{-\alpha}$, and $Y = \sum_{j\in\Phi_{\Bcal^{nc}}} g_j P r_j^{-\alpha}$, we have SIR outage probability $P_{\mathrm{out}} = \mathrm{Pr}(\Omega > 0)$, where $\Omega = \theta Y - X$ and $\theta$ is the predefined SIR threshold. 

\subsubsection{CGF of $\Omega$}
For homogeneous PPP, we know that the CGF of aggregate of impulse response positioned at each 2-D Poisson point is given by {$\Kcal(t) = 2 \pi \lambda \int (\Ebb_G[e^{-t g \ell(r)}] - 1) r \ud r$}, where $\ell(r)$ is the deterministic path-loss function and $g$ is the random channel gain. For our case, $\ell(r) = Pr^{-\alpha}$. Thus, $\Kcal_X(-t) = 2 \pi \lambda  \int_a^R (\Ebb_G[e^{t g P r^{-\alpha}}] - 1) r \ud r ]$ and $\Kcal_Y(\theta t) = 2 \pi \lambda \int_R^\infty (\Ebb_G[e^{- t \theta g P r^{-\alpha}}] - 1) r \ud r ] $. Consequently, we can write
\begin{align} 
\Kcal_\Omega(t) &= 2 \pi \lambda \big[ \int_a^R (\Ebb_G [e^{t g P r^{-\alpha}}] - 1) r \ud r + \int_R^\infty (\Ebb_G[e^{- t \theta g P r^{-\alpha}}] - 1) r \ud r \big],\nonumber\\
&= 2 \pi \lambda \big[ \int_a^R (\Mcal_G(-t P r^{-\alpha}) - 1) r \ud r  + \int_R^\infty (\Mcal_G(t \theta P r^{-\alpha}) - 1) r \ud r  \big]. 
\label{eqn:CGF-omega-rep-case}
\end{align}
Note that $\Ebb_G [e^{t g P r^{-\alpha}}] \equiv \Mcal_G(-t P r^{-\alpha})$ and $\Ebb_G [e^{-t \theta g P r^{-\alpha}}] \equiv \Mcal_G (t \theta P r^{-\alpha})$. Here the non-zero lower limit $a$ allows us to avoid the singularity at origin of the unbounded path-loss function $\ell(r)$. Also, it allows us to model the exclusion region. It is very important to set this parameter correctly, since it determines the heaviness of the tail of the resulting distributions. Small exclusion regions produce distributions with heavier tails while large exclusion regions produce distributions with lighter tails.

\begin{proposition}[Cumulants for $\Omega$]
 The $n$-th cumulant of $\Omega$ for our large scale cellular network model is given by
\beq 
\kappa_n(\Omega) = \kappa_n^{\mathrm{lim}} (\Omega) [1 + ((-\theta)^n - 1) u^{-n\alpha + 2}], 
\eeq
where $u = R/a$ and $ \kappa_n^{\mathrm{lim}} (\Omega) = ( -1)^n \frac{2 \pi  \lambda \mu_n(G) P^n}{n\alpha-2} a^{-n\alpha+2}$.
\label{prop:cumulant-rep-model}
 \end{proposition}
 
 \begin{IEEEproof}
See \textbf{Appendix A}.
 \end{IEEEproof}
 
 \begin{corollary} $\lim_{u \rightarrow \infty} \kappa_n(\Omega) = \kappa_n^{\mathrm{lim}}(\Omega). $ 
 \label{coro:lim-cumulant-rep-model}
 \end{corollary}
 
 \begin{corollary} Assuming $a<<R$, so that $\kappa_n(\Omega) \sim \kappa_n^{\mathrm{lim}}(\Omega)$, the skewness squared and excess kurtosis\footnote{The skewness and excess kurtosis of a random variable $X$ is defined as $\mathrm{Skew}(X) = \frac{\kappa_3(X)}{\kappa_2(X)^{3/2}}$  and $\mathrm{Ex.\,Kurt}(X) = \frac{\kappa_4(X)}{\kappa_2(X)^2}$.} of $\Omega$ are
 $\mathrm{Skew}^2(\Omega) = \frac{1}{2\pi\lambda a^2}\frac{(\alpha-1)^3}{(3\alpha-2)^2} \mathrm{Skew}^2(G)$ and $\mathrm{Ex.\,Kurt}(\Omega) = \frac{1}{\pi\lambda a^2}\frac{(\alpha-1)^2}{2\alpha-1} \mathrm{Ex.\, Kurt}(G).$  
 
 \label{coro:skew-kurtosis-rep-model}
 \end{corollary}
 \tbf{Remark:} According to Corollary \ref{coro:lim-cumulant-rep-model}, when $a << R$, the cumulants are independent of threshold $\theta$. Also, according to Corollary \ref{coro:skew-kurtosis-rep-model}, without losing much generality, we can see that both the skewness as well as kurtosis of $\Omega$ decreases as $a$ and $\lambda$ increases. This implies that the Gaussian approximation of $\Omega$ is valid only for large $a$ and $\lambda$.

  
\section{SIR Outage Calculation Using Cumulants/Moments}
As mentioned in Section~I, the moment problem can be analytically solved via a number of equivalent methods, such as by using continued fractions, rational functions, or orthogonal polynomials. These three concepts are intimately related to each other (see \cite{Akhiezer1965,Szego1975} for details). The rational functions were exploited in \cite{Amindavar1994a,Amindavar1994b,Stokes1998}. 

{\bf We want to stress that it is easier to calculate the cumulants of $\Omega$ than its moments} as mentioned in Section~III. Also, for the case with stochastic geometry, the cumulants can be found using Campbell's formula as given in Proposition \ref{prop:cumulant-rep-model}. Thus, in order to use moment-based techniques, we first recommend the calculation of cumulants and then its conversion into moments using the Bell polynomials. 

In this section, we will briefly overview the moment problem and the method of reconstructing a PDF from the cumulants/moments using the orthogonal polynomials. The choice of weighting functions  for orthogonal polynomials is then discussed. Finally, we detail the  outage evaluation using orthogonal polynomials,  with Gaussian and Student's-$t$ taken as base distributions.

\subsection{The Moment Problem}
The moment problem has long been studied by mathematicians. \textbf{Curiously, these techniques have not been comprehensively exploited for outage evaluations in wireless networks.} The probabilistic moment problem can be described as follows: let a sequence $\{\mu_i, i=0,1,\ldots\}$ of real numbers be given. Find the probability distribution on $\Ical \subseteq \Rbb$ such that $\mu_i = \int_\Ical x^i \ud F(x)$ for $i = 0,1,\ldots$.  There are three important questions related to the moment problem: (i) Does the distribution $F$ exist? (ii) If $F$ exists, is it uniquely determined by the moments $\{\mu_i\}$? (iii) How is the distribution reconstructed? If there is a unique distribution for the given moments, then the distribution is said to be \tit{determinate}. Otherwise it is said to be \tit{indeterminate}. The moment problems are classified according to the support of the distribution, i.e., when the two end-points are finite, one end-point is finite, or no end-point is finite, as summarized in Table \ref{table:types-of-moment-problems}. 

\begin{table}[ht]
\footnotesize
\centering
\caption{Types of Moment Problems}
     \begin{tabular}{lcl}
     \hline
        \hline
\tbf{Moment Problem} & \tbf{Support} & \tbf{Determinancy} \\
\hline
Hausdorff & compact & always determinant  \\
Stieltjes & $\Rbb_+$ & not always determinant  \\
Hamburger & $\Rbb$  & not always determinant  \\
\hline
\hline
\end{tabular}
\label{table:types-of-moment-problems}
\end{table}
\normalsize

It is important to be able to distinguish the type of moment problem at hand, and hence construct appropriate solution. Not being able to make the distinction can lead to poor or even incorrect solution. Our outage problem clearly belongs to the Hamburger moment problem (see Table \ref{table:types-of-moment-problems}) since $\Omega \in \Rbb$. Also,  by the nature of our problem, we can assume that a distribution corresponding to the moments exists. 
In the following, we will briefly discuss the reconstruction of a PDF, when its moments are given, using orthogonal polynomials.


 \subsection{Charlier Expansion}
 Given the moments $\mu_n(X)$ for $n = 0,1,2,\ldots,$ of a random variable $X$, let the reconstructed PDF be given by
 \beq
 f_X(x) = w(x) \sum_{k=0}^\infty a_k \phi_k(x),
 \label{eqn:pdf-reconst-1}
 \eeq
 where $w(x)$ is an initial density approximant (or weight function). The $\phi_k(x)$ are orthogonal polynomials associated with $w(x)$, such that the orthogonality condition 
$\int \phi_m(x) \phi_n(x) w(x) \ud x = C_n \delta_{mn}$ is satisfied,
where $\delta_{mn}$ is the Kronecker delta and $C_n = \int [\phi_n(x)]^2 w(x) \ud x$ is a normalizing constant. 
Using the orthogonality condition, the value of $a_k$ can be recovered by multiplying both sides of (\ref{eqn:pdf-reconst-1}) by $\phi_k(x)$ and integrating with respect to $x$, such that
 \beq
 a_k = \frac{1}{C_k}\int \phi_k(x) f_X(x) \ud x.
 \label{eqn:orthogonal-moment}
 \eeq
 The $a_k$s are also known as \tit{orthogonal moments}. Let the orthonormal polynomial $\phi_k(x)$ be given by $\phi_k(x) = \sum_{i=0}^k \phi_{ki} x^i$.
 Substituting this expression for $\phi_k(x)$ in (\ref{eqn:orthogonal-moment}), we  obtain
$ a_k = \frac{1}{C_k} \int \sum_{i=0}^k \phi_{ki} x^i f_X(x) \ud x = \frac{1}{C_k} \sum_{i=0}^k \phi_{ki} \int x^i f_X(x) \ud x.$
 Therefore, we have
 \beq 
 a_k = \frac{1}{C_k}\sum_{i=0}^k \phi_{ki} \mu_i(X). 
 \label{eqn:orthogonal-moment-1}
 \eeq
 Hence, we have the reconstructed PDF as
 \beq
 f_X(x) = w(x) \sum_{k=0}^\infty \frac{1}{C_k} \left(\sum_{i=0}^k \phi_{ki} \mu_i(X) \right)  \phi_k(x).
 \eeq
Such expansions are known as \tit{Charlier expansion}. For existence, uniqueness, and convergence of such series, the readers are referred to \cite{Szego1975}. In general, the series converges in least square sense if $f_X(x) \in \Lbb^2(w,\Ical)$ (see \cite[Ch. 2]{Szego1975}). Likewise, the CDF can be found by integrating (\ref{eqn:pdf-reconst-1}).

 \subsection{Choice of Weighting Functions and Associated Orthogonal Polynomials}
As mentioned earlier, the solution to the moment problem can be found by using appropriate orthogonal polynomials for a given weight function. The weight functions of the three classical orthogonal polynomials (Jacobi, Laguerre, and Hermite polynomials~\cite{Olver2010,Szego1975}), correspond to the initial approximant densities (Beta, Gamma, Gaussian) for particular moment problems (Hausdorff, Stieltjes, Hamburger), as summarized in Table \ref{table:classical-ortho-polys}.
 \begin{table}[ht]
\footnotesize
\centering
\caption{Weight functions and their associated orthogonal polynomials}
     \begin{tabular}{lllll}
     \hline
        \hline
\tbf{Weight function}, $\mathbf{w(x)\propto}$ & \tbf{Support} & \tbf{Distribution Name} & \tbf{Moment Problem} & \textbf{Associated Polynomial}
\\
\hline
 $(1-x)^\alpha(1+x)^\beta$ & $-1 \leq x \leq 1$ &  Beta  & Hausdorff  & Jacobi \\
$x^\alpha e^{-x}$ & $x \geq 0$ & Gamma  & Stieltjes  & Laguerre  \\
 $e^{-x^2/2}$ & $x \in \Rbb$  & Gaussian  & Hamburger & Hermite  \\
 $\left(1+\frac{x^2}{\nu} \right)^{-\frac{\nu+1}{2}}$ & $x \in \Rbb$ & Student's $t$ & Hamburger & Krishnamoorthy \\
 $\exp\left( \int \frac{a_1 x + a_0}{b_2 x^2 + b_1 x + b_0} \,\ud x \right)$ & - & Pearson family & - & Romanovsky-Hildebrant  \\
\hline
\hline
\end{tabular}
\label{table:classical-ortho-polys}
\end{table}
\normalsize 
Apart from these classical weights, we can use other non-classical weights and their corresponding non-classical orthogonal polynomials, which can be generated through \tit{Gram-Schmidt process}. In general, closer the initial approximating distribution to the true distribution, the better the approximation. 

One possibility is to fit the first four moments to Pearson family of distributions (e.g., Beta, Student's-$t$, Gamma, and so on) as the initial approximant.  
The associated orthogonal polynomials for the Pearson family were systematically investigated by Romanovsky \cite{Romanovsky1924, Romanovsky1929} and Hildebrant \cite{Hildebrandt1931} (see \cite{DiaconisZabell1991} for general introduction). The special case for $t$-distribution was further studied by Krishnamoorthy \cite{Krishnamoorthy1951}. 
In general, a Pearson distribution is any solution to the \tit{Pearson differential equation}: $\frac{1}{w} \frac{\ud w}{\ud x} = \frac{a_0 + a_1 x}{b_0 + b_1 x + b_2 x^2}$. The orthogonal polynomials associated with a Pearson distribution is given by the following recurrence relation due to Hildebrant \cite{Hildebrandt1931}:
\beq 
\phi_{n+1}(x) + \left[n D'(x) - N(x)\right] \phi_n (x) + n \left[\frac{(n-1)}{2!} D''(x) - N'(x) \right] D(x) \phi_{n-1}(x) = 0,  
\label{eqn:hildebrant-recurrence}
\eeq 
where $N(x) = a_0 + a_1 x$ and $D(x) = b_0 + b_1 x + b_2 x^2$. The initial polynomials being $\phi_0(x) = 1$ and $\phi_1(x) = N(x)$. 

For the Hamburger problem, Pearson Type IV distribution is a suitable initial approximant with the entire real line as its support and which can be fitted to the first four moments. The PDF of Pearson Type IV distribution is given by $w(x) \propto \left(1+\frac{x^2}{\alpha^2} \right)^{-m} e^{-\beta \arctan(\frac{x}{\alpha})}$, which corresponds to the differential equation $\frac{1}{w} \frac{\ud w}{\ud x} =  - \frac{\alpha \beta + 2mx}{\alpha^2 + x^2}$. Here, the parameter $\beta$ is used to control the skewness of the distribution. If $\beta = 0$, then we have Student's $t$-distribution; and if $\beta=0$ and $m \rightarrow \infty$, we have Gaussian distribution. While the Type IV distribution is of considerable generality, in the following sections, for the sake of simplicity, we will consider only the Gaussian distribution and the $t$-distribution for the Hamburger problem.

\begin{table}[ht]
\footnotesize
\centering
\caption{Hermite and Krishnamoorthy polynomials}
     \begin{tabular}{l c c m{4cm} c c c}
     \hline
        \hline
\tbf{Name} & \tbf{Symbol}  & \tbf{Pearson D.E.} & \multicolumn{1}{c}{\tbf{Recurrence}} & \multicolumn{2}{c}{\tbf{Initializations}} & \tbf{Normalization} \\
  & $\mbf{\phi_n(x)}$ & $\mbf{\frac{1}{w} \frac{\ud w}{\ud x}}$  & \multicolumn{1}{c}{$\mbf{\phi_{n+1}(x)}$} & $\mbf{\phi_0(x)}$ & $\mbf{\phi_1(x)}$ & $\mbf{C_n}$ \\ 
\hline
 Hermite  & $He_n(x)$  &  $-x$  &  $x He_n(x) - n He_{n-1}(x)$ & $1$ & $x$ & $n!$ \\
 Krishnamoorthy & $T_n(x)$  & $-\frac{(v+1)x}{v + x^2}$ & $(n+v+1) x T_n(x) - n(n+v)(x^2+v)T_{n-1}(x)$ & $1$ & $(v+1)x$ & *  \\
\hline
\multicolumn{7}{l}{* For Krishnamoorthy polynomials, $C_n = \frac{2^{1-v+2n}\pi \sqrt{v}}{v-2n} \frac{\Gamma(n+1)\Gamma(v-n+1)}{\Gamma^2(\frac{v+1}{2}-n)}$} \\
\hline
\hline
\end{tabular}
\label{table:He-Kr-polys}
\end{table}
\normalsize
As noted in Table \ref{table:classical-ortho-polys}, the orthogonal polynomials associated with the standard normal distribution are the Hermite polynomials, while that of the $t$-distribution are the Krishnamoorthy polynomials. The $t$-distribution allows us to account for the large positive kurtosis; and hence it can serve as an initial approximant for heavy tailed distributions. Table \ref{table:He-Kr-polys} summarizes some of the properties of these polynomials. The readers are referred to \cite{Olver2010,Szego1975} for details on Hermite polynomials and \cite{Krishnamoorthy1951} for Krishnamoorthy polynomials.

\subsection{SIR Outage Evaluation Using Cumulants/Moments}
Given the cumulants and in turn the moments of $\Omega$ (via Bell polynomial) and considering the base distributions, i.e., Gaussian and Student's $t$,
we have derived the following results on SIR outage probability with Hermite and Krishnamoorthy polynomials, respectively.
\begin{proposition}[SIR Outage Probability with Hermite Polynomials]
Given the moments of $\Omega$, $\mu_n(\Omega)$ for $n = 1, 2, \ldots,$ if we assume the base distribution to be standard normal, then the SIR outage is given by 
\beq 
P_{out} = 1 - \frac{1}{\sqrt{2 \pi}} \sum_{k=0}^\infty a_k \sum_{i=0}^k (-1)^i 2^{\frac{i-1}{2}} \Gamma\left(\frac{i+1}{2}\right) h_{ki}, 
\label{eqn:outage-hermite}
\eeq
where $h_{ki}$ are coefficients of $k$-th order Hermite polynomial, $He_k(\omega) = \sum_{i=0}^k h_{ki} \omega^i$, and $a_k = \frac{1}{k!} \sum_{i=0}^k h_{ki}\mu_i(\Omega)$.
\label{prop:outage-hermite}
\end{proposition}

\begin{IEEEproof}
See \textbf{Appendix~B}.
\end{IEEEproof}

\begin{proposition}[SIR Outage Probability with Krishnamoorthy Polynomials]
Given the moments of $\Omega$, $\mu_n(\Omega)$ for $n = 1, 2, \ldots,$ if the base distribution is assumed to be Student's $t$, then the SIR outage is given by 
\beq 
P_{out} =  1 - \sum_{k=0}^{\lfloor v/2 \rfloor} a_k \sum_{i=0}^k (-1)^i \frac{v^{(i+1)/2}}{2} B\left(\frac{v - i}{2},\frac{1 + i}{2}\right) t_{ki}, 
\label{eqn:outage-krishnamoorty}
\eeq
where $t_{ki}$ are coefficients of $k$-th order Krishnamoorthy polynomial, $T_k(\omega) = \sum_{i=0}^k t_{ki} \omega^i$ and $a_k = \frac{1}{C_k}\sum_{i=0}^k t_{ki}\mu_i(\Omega)$. Assigning the value of $v$ using moment matching, $v = \frac{6}{\mathrm{Ex. \, Kurt}(\Omega)} + 4 $,
where $\mathrm{Ex.\, Kurt}(\Omega) =  \kappa_4(\Omega)/\kappa_2(\Omega)^2$. 
\label{prop:outage-krishnamoorty}
\end{proposition}
 
 \begin{IEEEproof}
See \textbf{Appendix C}.
 \end{IEEEproof}
      
 \tbf{Remark:} Note that expression using Krishnamoorthy polynomial is a finite sum, since for t-distribution, the moment of order $v$ or higher does not exist. Thus, the set of orthogonal polynomials associated with t-distribution is also finite. In fact, there are only $\left\lfloor \frac{v}{2} \right\rfloor$ orthogonal Krishnamoorthy polynomials, for a given parameter $v$ of the t-distribution.


 \section{Saddle Point Method using CGF}
Apart from the moments of $X$, if the CGF of $X$ also exists, then we can exploit a more powerful technique known as the \tit{saddle point approximation} (SPA) to compute the CDF of the random variable (see \cite{Butler2007} for general introduction). The saddle point method  serves as a compromise between the purely analytical and purely numerical approaches. In this approach, we need to evaluate a single point of the function to be integrated, called the \tit{saddle point}, based on which a semi-analytical formula for outage can be obtained. 

Note that the Gil-Pelaez inversion formula can be represented in terms of CGF as: 
\[ Q_{X}(x) = \frac{1}{2\pi \jmath} \int_{c- \jmath \infty}^{c+\jmath\infty} e^{\Kcal_X(t) - t x} \frac{\ud t}{t}, \]
where $c>0$ is a real constant lying in the convergence strip of $\Kcal_X(t)$. The dominant component of the integral is concentrated at the saddle point of $\Kcal_X(t) - tx$. The saddle point $\hat{t} = \hat{t}(x)$ is given by the solution of the saddle point equation $\Kcal'_X(\hat{t}) = x$. 
Now, suppose that $g$, $G$, and $\Lcal$ are the PDF, CDF, and CGF of the base distribution of $Z$, respectively, by which we want to approximate our target distribution. The dominant component of this base distribution is found at $\Lcal_Z(\breve{s}) - \breve{s}z$, where $\breve{s} = \breve{s}(z)$ is the saddle point root of $\Lcal'_Z(\breve{s}) = z$. After transforming the pair $(x, t) \mapsto (z, s)$ such that dominant components of these two distributions coincide, we obtain
 \beq 
 \Lcal_Z(\breve{s}) - \breve{s} z = \Kcal_X(\hat{t}) - \hat{t}x.
 \label{eqn:coincidence-CGF}
 \eeq
 
The task is to find an optimal choice of $\hat{z} = \hat{z}(x)$ from the above transformation process, when the right hand side is given. 
 \begin{figure}[h]
\begin{center}
	\includegraphics[scale=0.2]{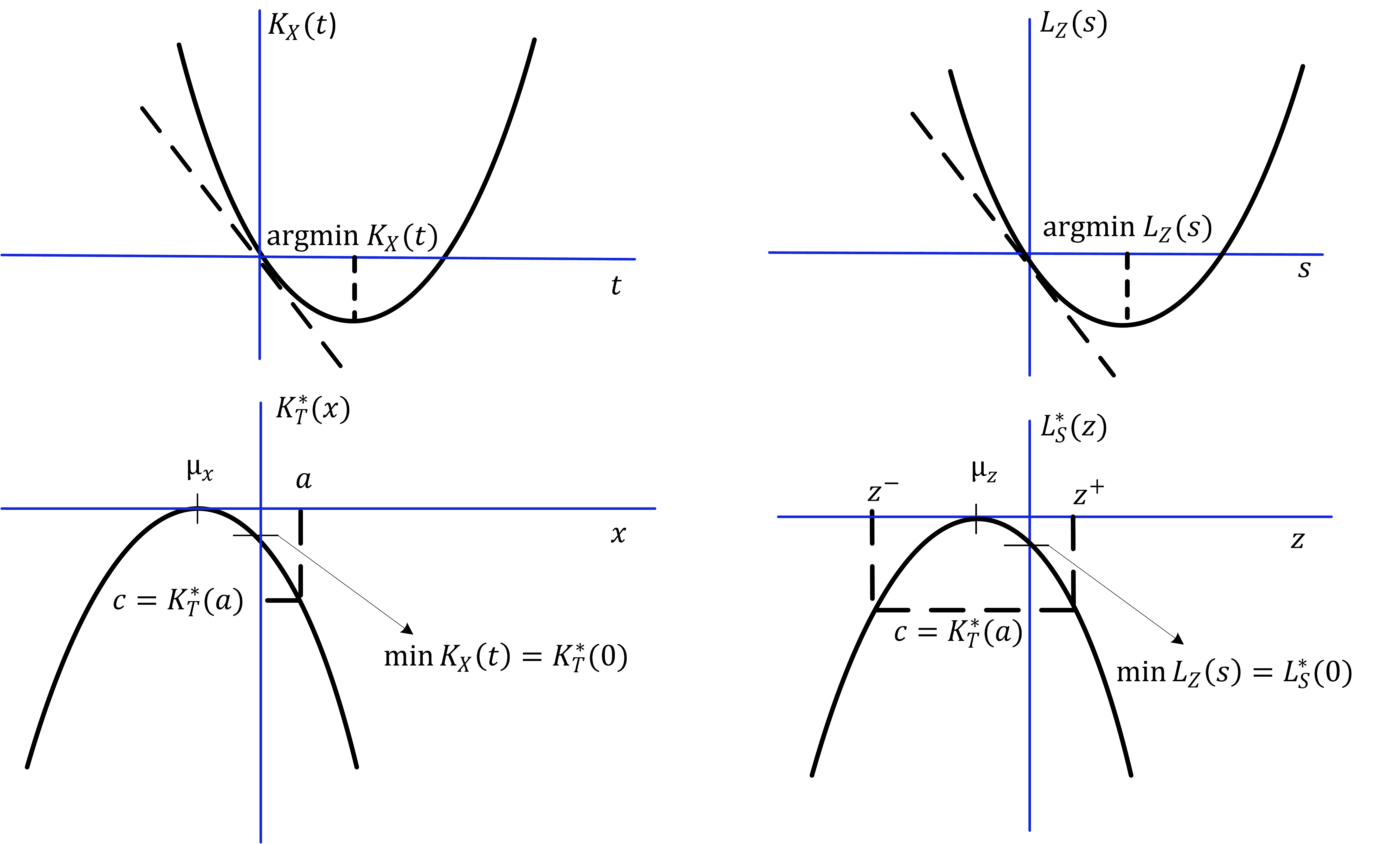}
	\caption{The Legendre-Fenchel duality. }
	\label{fig:legendre-fenchel-duality}
 \end{center}
\end{figure} 
Also note that the quantities $\Kcal_T^*(x) = \hat{t}x - \Kcal_X(\hat{t})$ and $\Lcal_S^*(z) = \breve{s}z - \Lcal_Z(\breve{s})$ appearing in the left-hand and right-hand of (\ref{eqn:coincidence-CGF}) are Legendre-Fenchel (LF) transforms of $\Kcal_X$ and $\Lcal_Z$, respectively, as shown in Fig. \ref{fig:legendre-fenchel-duality}. As per the definition of CGF, both $\Kcal_X(t)$ and $\Lcal_Z(s)$ are convex, whereas their LF transforms $\Kcal_T^*(x)$ and $\Lcal_S^*(z)$ are concave with respect to $x$ and $z$. By the property of LF transform, the maxima/minima of the dual function is given by the value of the intercept on the ordinate axis of the primal function, whereas the location of the maxima/minima of the dual function is given by the slope of the primal function at that intercept. Thus, $\Kcal_T^*(x)$ and $\Lcal_S^*(z)$ have unique maxima of zero at their means $x = \Ebb[X] = \Kcal'_X(0)$ and $z = \Ebb[Z] = \Lcal'_Z(0)$, respectively. These maxima corresponds to their dual variables $\hat{t}=0$ and $\breve{s} = 0$. Likewise, the minima of $\Kcal_X(t)$ and $\Lcal_Z(s)$ correspond to the ordinate intercept of their dual function, $\min \Kcal_X(t) = \Kcal_T^*(0)$ and $\min \Lcal_Z(t) = \Lcal_S^*(0)$. 

As such, because of the concavity, for a given value of $x$, there can be two possible optimal choices for $\hat{z}$ in (\ref{eqn:coincidence-CGF}). For the unique case when $x = \Ebb[X] = \Kcal'_X(0)$, there is only one possible choice of $\hat{z} = \Lcal_Z'(0)$ for the base distribution. For $x \neq \Ebb[X]$, there are two solutions $\hat{z}_{-}(x) < \Ebb[Z] < \hat{z}_{+}(x)$ on either side of the mean for the base distribution. The root of $z$ should be such that its relative position with respect to its mean $\Ebb[Z]$ should  match with the relative position of $x$ with respect to its mean $\Ebb[X]$. Thus,
\beq
\hat{z}(x) = \left\{ \begin{array}{lcr}
\hat{z}_{-}(x), & \mathrm{if} & x < \Ebb[X] \\
\Lcal'_Z(0), & \mathrm{if} & x = \Ebb[X] \\
\hat{z}_{+}(x), & \mathrm{if} & x > \Ebb[X]. 
  \end{array} \right.  
  \label{eqn:optimal-z}
\eeq
Subsequently, we have the following proposition by Wood, Booth, and Butler.  
\begin{proposition}[CDF Approximation using SPA \cite{WoodBoothButler1993}]
Suppose $X$ has a continuous distribution $F_X(x)$ with CGF $\Kcal_X(t)$. The $(g, G)$-based saddle point CDF approximation for $F_X(x)$ is 
 \beq 
 \hat{F}_X(x) = G_Z(\hat{z}) + g_Z(\hat{z}) \left[ \frac{1}{\hat{s}} - \frac{1}{\hat{u}} \right]
 \label{eqn:WBB-formula}
 \eeq 
 where $\hat{z}$ is given in (\ref{eqn:optimal-z}), $\hat{s} = \breve{s}(\hat{z})$ is the saddle point for $\hat{z}$ with respect to the base CGF and $\hat{u} = \hat{t}\sqrt{\frac{\Kcal''_X(\hat{t})}{\Lcal''_Z(\hat{s})}}$.
 \end{proposition}
 
This CDF approximation is  independent of the location and scale of the base distribution \cite{WoodBoothButler1993}. Also, given a base distribution, the authors recommend the moment matching method to derive the parameters of the base distribution, i.e., $ \Lcal_Z^{(n)}(\breve{s}) = \Kcal_X^{(n)}(\hat{t}),$ for $n=1,2,\ldots.$ 
 
Table \ref{table:BWW-for-various-distr} summarizes some popular choices for base distributions that are characterized by their CGFs. The first of these, with standard normal base, gives the famous \tit{Lugannani-Rice formula} \cite{LugannaniRice1980}, whereas the remaining two base distributions have been discussed in \cite{WoodBoothButler1993}. Starting with $(x,\hat{t})$, we need to find $\hat{z}$, $\hat{s}$, and $\hat{u}$ before the calculation of $\hat{F}_X(x)$. The calculation can be performed in the following sequence: 
\begin{equation}
(x,\hat{t}) \rightarrow \mathrm{parameters\; of}\; G_Z \rightarrow \hat{z}(x) \rightarrow \hat{s} \rightarrow \Lcal''_Z(\hat{s}) \rightarrow \hat{u} \rightarrow \hat{F}_X(x).
\end{equation}

 \begin{table}[ht]
\footnotesize
\centering
\caption{Some base distributions and their saddle points}
\begin{tabular}{*{5}{l}}
 \hline
 \hline
\tbf{Base distribution} & $\mathbf{\Lcal_Z(s)}$ & $\mathbf{\breve{s}(z)}$ & $\mathbf{\hat{z}(x)}$ & $\mathbf{\hat{u}}$ \\
\hline
 Standard Normal & $\frac{s^2}{2}$ &  $z$  & $\mathrm{sgn}(\hat{t}) \sqrt{2(\Kcal'_X(\hat{t}) - \hat{t}x)}$  & $\hat{t} \sqrt{\Kcal''_X(\hat{t})}$ \\
 Chi-square & $- \frac{\alpha}{2} \log(1-2s)$ & $\frac{1}{2}(1-\frac{\alpha}{z})$ & $-\alpha W(- \exp(-\frac{2c}{\alpha} -1))$ & $\frac{\hat{t}}{\hat{z}} \sqrt{\frac{\Kcal''_X(\hat{t})}{2}}$ \\
 Inverse Gaussian & $\mu^{-1} - (\mu^{-2} - 2 s)^{1/2}$ & $\frac{1}{2}(\mu^{-2} - z^{-2})$ & $\mu + \mu^2(c + \mathrm{sgn}(\hat{t})\sqrt{c^2 + 2 c \mu^{-1}})$ & $\frac{\hat{t}\sqrt{\Kcal''_X(\hat{t})}}{\hat{z}^{3/2}}$ \\
\hline
\multicolumn{5}{l}{\tbf{Note:} $\alpha = 8 \eta^{-1}$ and  $\mu = \frac{\eta}{9}\left(1 + \mathrm{sgn}(\hat{t})\sqrt{\frac{-2\eta c}{9}}\right)^{-1}$ where $\eta = \frac{\Kcal_X'''(\hat{t})^2}{\Kcal_X''(\hat{t})^3}$; $c = \Kcal^*_T(x)$} and $W(\cdot)$ is Lambert's W function\\
\hline
\hline
\end{tabular}
\label{table:BWW-for-various-distr}
\end{table}
\normalsize

Now we introduce a relatively new,  four parameter distribution known as the \tit{normal-inverse Gaussian} (NIG) distribution, which is a special case of more general hyperbolic distributions \cite{Nielsen1978}, to allow us more flexibility in adjusting the skewness as well as the kurtosis. The parameters can be explicitly solved for given cumulants during the moment matching process \cite{Eriksson2004}. Since the distribution is defined on the entire real line, it is suitable for our outage problem. 
 The PDF of NIG distribution is 
 \[ f_X(x; \alpha, \beta, \mu, \delta) = \frac{\alpha}{\pi \delta}\exp(\delta \sqrt{\alpha^2 - \beta^2} + \beta(x-\mu)) \frac{K_1 \left(\alpha \delta \sqrt{1+(\frac{x-\mu}{\delta})^2}\right)}{\sqrt{1+(\frac{x-\mu}{\delta})^2}}, \]
 where $x\in\Rbb$, $\alpha>0$, $\delta>0$, $\mu\in\Rbb$, $0<|\beta|<\alpha$. The $K_1(\cdot)$ is modified Bessel function of second kind with index 1. The CGF of the distribution is 
 \[\Kcal_X(t) = \mu t + \delta [\sqrt{\alpha^2 - \beta^2} - \sqrt{\alpha^2 - (\beta + t)^2}].\]

The process that gives rise to the NIG distribution is as follows: If $X$ is normal distributed when conditioned on $Y$, with mean $\mu + \beta Y$ and variance $Y$, so that $f_{X|Y}(x|y) = N(\mu+\beta Y, Y)$, and $Y$ itself follows an inverse Gaussian distribution $f_Y(y) = IG(\delta,\sqrt{\alpha^2 - \beta^2})$, then the unconditional distribution of $X$ is said to be normal-inverse Gaussian $f_X(x) = NIG(\alpha,\beta,\mu,\delta)$. The NIG distribution tends to Gaussian distribution as $\alpha \rightarrow \infty$.

\begin{table}[ht]
\footnotesize
\centering
\caption{SPA using Normal-Inverse Gaussian base distribution}
\begin{tabular}{ll}
 \hline
 \hline
\multicolumn{1}{c}{\tbf{Saddle point parameters}} & \multicolumn{1}{c}{\tbf{NIG parameters}}  \\
\hline
$\Lcal_Z(s) = \sqrt{\alpha^2 - \beta^2} - \sqrt{\alpha^2 - (\beta + s)^2}$  &  $\mu = 0$ \\
$\breve{s}(z) = -\beta + \frac{\alpha z}{\sqrt{1+ z^2}}$ & $\delta = 1$ \\
$\hat{z}(x) = \frac{d \beta + \sgn(\hat{t}) \alpha \sqrt{d^2 + (\beta^2 -1)}}{\alpha^2 - \beta^2}$  & $\alpha = 9 [(3\rho - 5 \eta)(3\rho - 4\eta)]^{-1/2}$\\
$L''_Z(\breve{s}) = \frac{z^3 + z}{\breve{s}+\beta}$ & $\beta = \frac{e \hat{z} + \sgn(K'''(t)) \sqrt{e^2 \hat{z}^2 - (e^2-\alpha^2)(1-\hat{z}^2)}}{1+\hat{z}^2}$ \\
\hline
\hline
\end{tabular}
\label{table:BWW-for-NIG-distr}
\end{table}
\normalsize
 
In Table \ref{table:BWW-for-NIG-distr}, we have summarized the saddle point method using NIG distribution. Here we take advantage of the fact that (\ref{eqn:WBB-formula}) is independent of location and scaling. The skewness and excess kurtosis are represented by $\eta = \frac{\Kcal'''(\hat{t})^2}{\Kcal''(\hat{t})^3}$ and $\rho = \frac{\Kcal^{\textrm{\romannumeral 4}}(\hat{t})}{\Kcal''(\hat{t})^2}$. Here $c = \Kcal_X(\hat{t}) + x \hat{t}$, $d = \sqrt{\alpha^2 - \beta^2} - c$, and $e = c + \alpha\sqrt{1+z^2}$. Also, we have the relation $\hat{z} = \sgn(\hat{t})(\frac{3\rho}{\eta} - 5)^{-1/2}$.

 
 \section{Application of SPA: Case Studies}
In this section, we consider approximating the SINR outage for all modeling scenarios listed in Section~I, using the saddle point methods explained in Section~V. To apply SPA, we require the first and second derivatives of CGF of $\Omega$ and the solution to the saddle point equation, $\Kcal_\Omega'(\hat{t}) = 0$. In the following, we provide the required derivatives and the exact closed-form expression of the saddle point, $\hat{t}$, whenever possible. When noise is neglected, the SIR outage probability can be computed as $P_{out} = Q_\Omega(0) = 1 - F_\Omega(0)$, where the $F_\Omega(0)$ is obtained from the Wood-Booth-Butler formula (\ref{eqn:WBB-formula}) $\hat{F}_\Omega(0) = G_Z(\hat{z}) + g_Z(\hat{z}) \left[ \frac{1}{\hat{s}} - \frac{1}{\hat{u}} \right]$. The parameters $\hat{z}$, $\hat{u}$, and $\hat{s}$ are given by Table \ref{table:BWW-for-various-distr} and \ref{table:BWW-for-NIG-distr}.

\subsection{Uncertainty due to Multi-Path Channel Fading and Number of Interferers}
In our generic model, consider the case where $\Omega = \theta Y - X$, $Y = \sum_{i=1}^N G_i$, and $X = \sum_{j=1}^M G_j$ where $G_i$ are IID random variables representing fast fading. Also, let the variables $M$ and $N$ be random. The CGFs of compound distribution is $\Kcal_Y(t) = \Kcal_N(-\Kcal_G(t))$ and $\Kcal_X(t) = \Kcal_M(-\Kcal_G(t))$, while $\Kcal_\Omega(t) = \Kcal_Y(\theta t) + \Kcal_X(-t)$.   In the following, we will consider the cases when $N$ and $M$ obeys Poisson and binomial distributions. In general, if $N \sim \mathrm{Poisson}(\lambda)$, then the  CGF of $N$ is $\Kcal_N(t) = \lambda(e^{-t}-1)$. Thus, $\Kcal_Y(t) = \Kcal_N(-\Kcal_G(t)) = \lambda(e^{\Kcal_G(t)}-1) = \lambda(\Mcal_G(t)-1)$. Similarly, if $N \sim \mathrm{Binomial}(L,p)$, then the CGF of $N$ is $\Kcal_N(t) = L \log(q + p e^{-t})$. Thus, $\Kcal_Y(t) = \Kcal_N(-\Kcal_G(t)) = L \log(q + p e^{\Kcal_G(t)}) = L \log(q + p \Mcal_G(t))$. 

\subsubsection{Nakagami-$m$ fading and Poisson aggregation}
For Nakagami-$m$ fading, the channel gain is given by the gamma distribution, $G_i \sim \mathrm{Gamma}(\alpha,\beta)$, such that its MGF is $\Mcal_G(t) = (1+ \frac{t}{\beta})^{-\alpha}$. Let the Poisson aggregation be given by $M \sim \mathrm{Poisson}(\lambda_1)$ and $N \sim \mathrm{Poisson}(\lambda_2)$. Using the relation for compound Poisson distribution, we have $\Kcal_Y(t) = \lambda_2[(1+\frac{t}{\beta})^{-t} - 1]$ and $\Kcal_X(t) = \lambda_1[(1+\frac{t}{\beta})^{-t} - 1]$. Thus, we have $\Kcal_\Omega(t) = \Kcal_Y(\theta t) + \Kcal_X(-t) = \lambda_2[(1+\frac{\theta t}{\beta})^{-t} - 1] + \lambda_1[(1 - \frac{t}{\beta})^{-t} - 1]$. Taking the derivative of $\Kcal_\Omega(t)$ with respect to $t$, and solving the saddle point equation $\Kcal'_\Omega(\hat{t}) = 0$, we can analytically obtain the saddle point, as stated below, by basic algebra.
\begin{proposition}
For Nakagami-$m$ fading and Poisson aggregation, the solution to $\Kcal'_\Omega(\hat{t}) = 0$ is $\hat{t} = \frac{\beta}{1+\theta}\left[1 - (\frac{\theta \lambda_2}{\lambda_1})^{-\frac{1}{\alpha + 1}}\right]$
and $\Kcal''_\Omega(t) = \frac{\alpha(\alpha+1)}{\beta^2}\left[ \lambda_2 \theta^2 (1+\frac{\theta t}{\beta})^{-\alpha-2} + \lambda_1 (1-\frac{t}{\beta})^{-\alpha-2} \right].$
\end{proposition}

\begin{corollary}
When $\lambda_1 = \lambda_2$, we have $\hat{t} = \frac{\beta(1-\theta^{1/(\alpha+1)})} {1+\theta}$.
\end{corollary}

\begin{corollary}
For Rayleigh fading, where $\alpha = 1$, $\hat{t} = \frac{\beta}{1+\theta}\left[1 - (\frac{\theta \lambda_2}{\lambda_1})^{-\frac{1}{2}}\right]$. Furthermore, if $\lambda_1 = \lambda_2$ too, then $\hat{t} = \frac{\beta(\sqrt{\theta}-1)}{\sqrt{\theta}(1+\theta)}$.
\end{corollary}


\tbf{Remark:} The parameters $\lambda_1$ and $\lambda_2$ can be interpreted as arising through a thinning process of a parent Poisson distribution with parameter $\lambda$ over a common spatial area, such that $\lambda_1 = p \lambda_1$ and $\lambda_2 = (1 - p) \lambda_2$, where $p$ can be interpreted as probability of cooperation. Alternatively, $\lambda_1$ and $\lambda_2$ can arise due to Poisson point process over two mutually exclusive spatial regions of differing areal sizes, as in our representative cellular network.

 \subsubsection{Nakagami-$m$ fading and binomial aggregation}
 Now, consider instead the case when we have $L$ total nodes such that $M \sim \mathrm{Binomial}(L,p)$ and $N = L - M \sim \mathrm{Binomial}(L,q)$, where $p+q = 1$. Here $p$ is interpreted as the probability of cooperation. As before, for Nakagami-$m$ fading, the channel gain is given by the gamma distribution, $G_i \sim \mathrm{Gamma}(\alpha, \beta)$, such that its MGF is $\Mcal_G(t) = (1+ \frac{t}{\beta})^{-\alpha}$. Using the relation for compound binomial distribution, we have $\Kcal_Y(t) = L \log(p + q (1 + \frac{t}{\beta})^{-\alpha})$ and $\Kcal_X(t) = L \log(q + p (1 + \frac{t}{\beta})^{-\alpha})$. Hence, we have $\Kcal_\Omega(t) = \Kcal_Y(\theta t) + \Kcal_X(-t) = L \log(p + q (1 + \frac{\theta t}{\beta})^{-\alpha}) + L \log(q + p (1 - \frac{t}{\beta})^{-\alpha})$. Taking the derivative of $\Kcal_\Omega(t)$ with respect to $t$, the saddle point equation $\Kcal'_\Omega(\hat{t}) = 0$ can be simplified.
 
 \begin{proposition}
 For Nakagami-$m$ fading and binomial aggregation, the solution to $\Kcal'_\Omega(\hat{t}) = 0$ is found by solving $p^2(1+\frac{\theta \hat{t}}{\beta})^{\alpha+1} + pq(1+\theta)\frac{\hat{t}}{\beta} - q^2(1-\frac{\hat{t}}{\beta})^{\alpha+1} = 0$. Also, 
 \begin{align*} \Kcal''_\Omega(t) =& \frac{L}{\beta^2}\Big[ -\frac{p^2 \alpha^2 (1 - t/\beta)^{-2 - 2\alpha}}{(q + p (1 - t/\beta)^{-\alpha})^2} 
+ \frac{p (\alpha + 1) \alpha (1 - t/\beta)^{-2 - \alpha}}{q + p (1 - t/\beta)^{-\alpha}} \\
& - \frac{q^2 \alpha^2 \theta^2 (1 + (t \theta)/\beta)^{-2 - 2\alpha}}{p + q (1 + (t \theta)/\beta)^{-\alpha})^2} 
+ \frac{q (\alpha + 1) \alpha \theta^2 (1 + (t \theta) /\beta)^{-2 - \alpha})}{(\beta^2 (p + q (1 + (t \theta)/\beta)^{-\alpha}))} \Big]. 
 \end{align*}
  \end{proposition}
Unfortunately, we cannot solve the saddle point equation analytically and we need to resort to some numerical root finding technique. 
 \begin{corollary}
 For the Rayleigh fading, where $\alpha = 1$, $\hat{t} = \frac{- 2 \beta \theta (1 - pq) + \sqrt{4 \beta^2 \theta^2 (1 - pq)^2 - 4 \beta^2 \theta (q^2 - \theta p^2 ) (\theta q - p)}}{2 \theta (\theta p^2 - q^2) }.$ 
  \end{corollary}
 
 
 \subsection{Uncertainty due to Aggregate Interference and Distance-based Attenuation}
 From our representative model, we have the CGF of $\Omega$ as given by (\ref{eqn:CGF-omega-rep-case}). Since no fading is assumed, the channel is  deterministic. Thus the CGF of $\Omega$ is simplified to 
 \beq 
 \Kcal_\Omega(t) = 2 \pi \lambda \left[ \int_a^R (e^{t P r^{-\alpha}} - 1) r \ud r + \int_R^\infty (e^{- t \theta P r^{-\alpha}} - 1) r \ud r \right], 
 \label{eqn:CGF-omega-rep-case-C}
 \eeq
 where the channel gain is normalized to unity. The integrals and the derivatives of (\ref{eqn:CGF-omega-rep-case-C}) can be evaluated using incomplete Gamma functions. To find the derivatives of $\Kcal_\Omega(t)$, we will first give the following proposition.
 
\begin{proposition}
If $\Kcal(t) = 2 \pi \lambda \int_a^b (e^{-t P r^{-\alpha}} - 1) r \ud r$, then its $n$-th derivative is 
\beq 
\Kcal^{(n)}(t) = (-1)^n \frac{2\pi\lambda}{\alpha} \frac{(tP)^{2/\alpha}}{t^n}  \left[ \Gamma\left(-\frac{2}{\alpha},tPb^{-\alpha}\right) - \Gamma\left(-\frac{2}{\alpha},tPa^{-\alpha}\right) \right]. 
\label{eqn:n-th-derivative-CGF-rep-case-C}
\eeq
\label{prop:n-th-derivative-CGF-rep-case-C}
\end{proposition}

\begin{IEEEproof}
See \textbf{Appendix D}.
 \end{IEEEproof}

Since we have $\Kcal_\Omega(t) = \Kcal_Y(\theta t) + \Kcal_X(-t)$, the derivatives of $\Kcal_\Omega(t)$ immediately follows by applying (\ref{eqn:n-th-derivative-CGF-rep-case-C}) as given in the proposition below. 
\begin{proposition}
The $n$-th derivative of $\Kcal_\Omega(t)$ is $\Kcal^{(n)}_\Omega(t) = \Kcal^{(n)}_Y(\theta t) + \Kcal^{(n)}(-t)$, where 
\begin{align*}
\Kcal^{(n)}_Y(\theta t) &= (-1)^n \frac{2\pi\lambda}{\alpha} \frac{(\theta tP)^{2/\alpha}}{t^n} \gamma\left(-\frac{2}{\alpha}, t\theta PR^{-\alpha}\right), \\
\Kcal^{(n)}_X(-t) &= (-1)^n \frac{2\pi\lambda}{\alpha} \frac{(-tP)^{2/\alpha}}{t^n}  \left[ \Gamma\left(-\frac{2}{\alpha}, -tPR^{-\alpha}\right) - \Gamma\left(-\frac{2}{\alpha}, -tPa^{-\alpha}\right) \right], 
\end{align*}
where $\gamma(a,z)$ is the lower incomplete Gamma function, such that $\gamma(a,z)+\Gamma(a,z) = \Gamma(a)$.
\end{proposition}

\begin{IEEEproof}
By applying (\ref{eqn:n-th-derivative-CGF-rep-case-C}) of Proposition \ref{prop:n-th-derivative-CGF-rep-case-C} to $\Kcal_Y(\theta t)$ and $\Kcal_X(-t)$.
\end{IEEEproof}

Since $\Gamma(a,-z)$ is in general a complex number, we have to be careful when interpreting this result. To solve the saddle point equation $\Kcal'_\Omega(t) = 0$, we need to resort to numerical root finding technique such as the Newton-Raphson method. Since $\Kcal_\Omega(t)$ is convex by definition, the saddle point is essentially the unique global minima of the CGF. As such, we can also use numerical optimization techniques to find the saddle point.


 \subsection{Uncertainty due to Multi-path Channel Fading and Number of Interferers}
 This is the most difficult problem in the group. For this case, the CGF is given by (\ref{eqn:CGF-omega-rep-case}). Depending on the kind of fading channel assumed, the MGF $\Mcal_G$ may or may not exist. If $\Mcal_G$ exists, then the problem may be tackled by the usual saddle point method. Symbolically, the $n$-th derivative of (\ref{eqn:CGF-omega-rep-case}) is given by
\[ \Kcal^{(n)}_\Omega(t) = 2 \pi \lambda P^n \left[ (-1)^n \int_a^R \Mcal^{(n)}_G(-t P r^{-\alpha})  r^{-n\alpha+1} \ud r  + \theta^n \int_R^\infty \Mcal^{(n)}_G(t \theta P r^{-\alpha}) r^{-n\alpha+1} \ud r  \right]. \]
However, the closed-forms of the integral for the CGF and its derivatives may not be available, or be available in terms of special functions such as hypergeometric functions, Meijer's-$G$ functions, or Fox-$H$ functions. In general, numerical integration may be unavoidable when applying the SPA technique. However, since the Campbell's theorem allows us to calculate the cumulants easily, we can  find the SINR outage from the cumulants. This latter method can be adopted for fading distributions for which the $\Mcal_G$ does not exist.
 

\section{Numerical Results}

In this section, we will describe some of the numerical results that compare the outage probability obtained via Gil-Pelaez and saddle point method. For the saddle point method, we use the normal distribution as the base distribution, and hence the Lugannani-Rice formula.

\subsection{Uncertainty due to Fading and Number of Interferers}
\begin{figure}[h]
\begin{center}
	\includegraphics[width=4in]{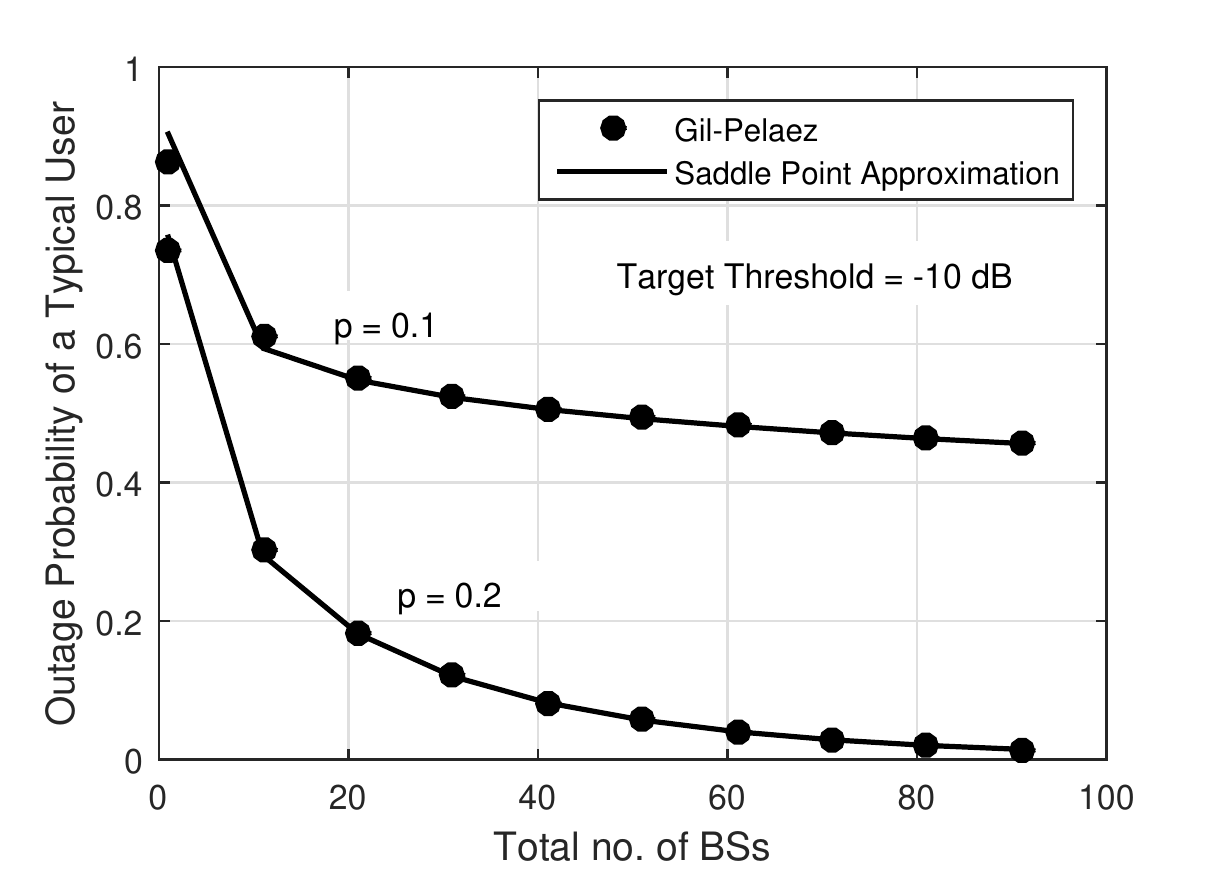}
	\caption{Case A: SIR outage vs. number of BSs at $\theta = 0$ dB, $P = 0$ dB}
	\label{fig:caseB-intensity}
 \end{center}
\end{figure}

For the case of binomial aggregation with Rayleigh fading, in Fig. \ref{fig:caseB-intensity}, we plot the outage probability as the total number of BSs is varied. In this figure, the SIR threshold is maintained at -10 dB, the transmit power of all BSs is 0 dB, while the probability of cooperation is assumed to be $0.1$ and $0.2$. Overall, we see that as the number of BSs increases, the outage tends to decrease and start to saturate at some level. The outage decreases faster when $p=0.2$ than when $p=0.1$, indicating that even a small change in the probability of cooperation leads to large gain in performance, especially when the number of BSs is large. We also observe that both the Gil-Pelaez formula and saddle point approximation (SPA) give very similar results.

\begin{figure}[h]
\begin{center}
	\includegraphics[width=4in]{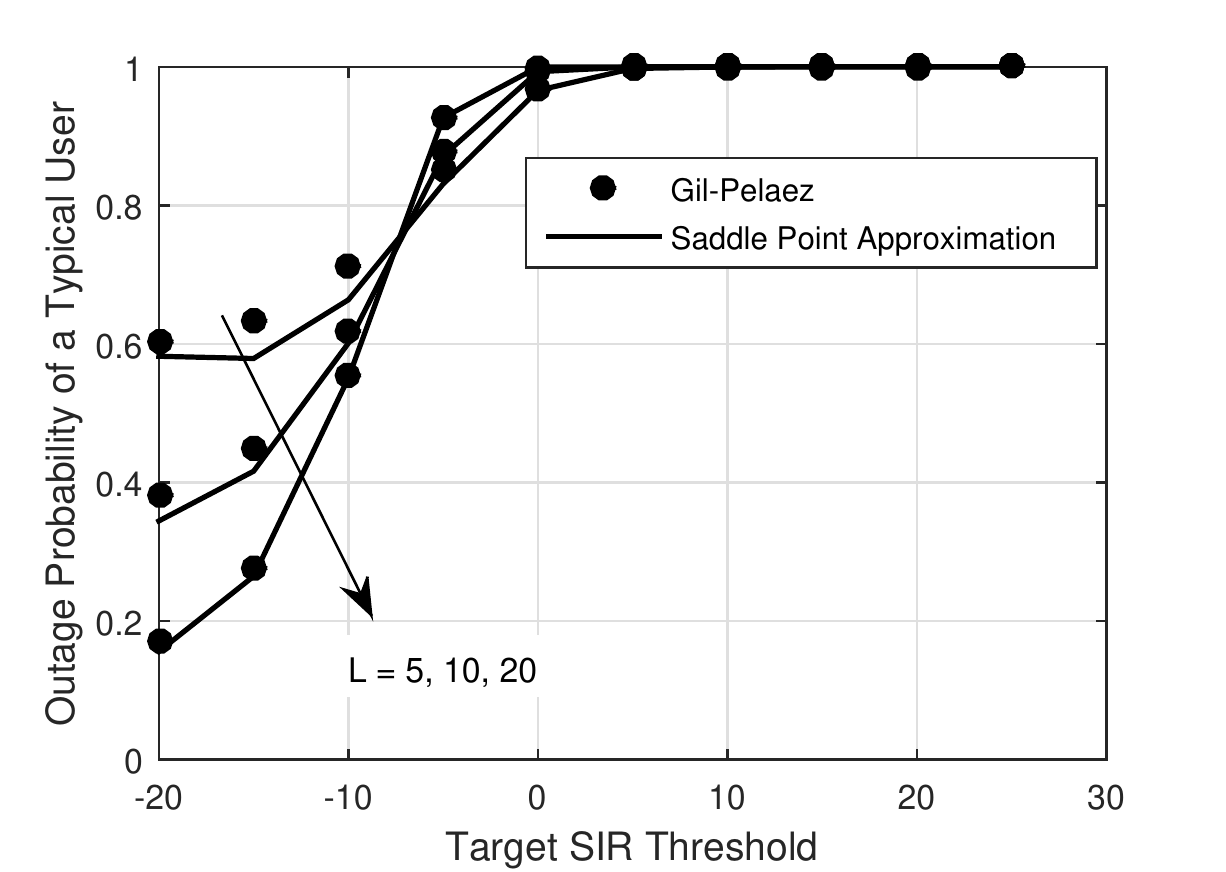}
	\caption{Case A: SIR outage vs. target threshold at $P = 0$ dB. }
	\label{fig:caseB-target}
 \end{center}
\end{figure}

Likewise, in Fig. \ref{fig:caseB-target}, we  plot the outage probability against SIR threshold, as the total number of BSs $L$ is varied as $5$, $10$, and $20$. Here the value of probability of cooperation is assumed to be $p=0.1$. As an overall trend, we see that as the threshold increase, so does the outage. At lower threshold levels, a typical user experiences lower outage with higher number of BSs, which is consistent with the conclusion derived from Fig. \ref{fig:caseB-intensity}. Beyond certain threshold, the case with higher BSs tends to suffer higher outage, indicating the dominance of interference. When we compare the results given by Gil-Pelaez and SPA, we notice that the SPA gives similar predictions as Gil-Pelaez when $L$ is higher. For lower $L$, the SPA starts to lose its accuracy at lower threshold. This can be explained by the loss of ``Gaussianity" at lower $L$. 

\subsection{Uncertainty due to Aggregate Interference and Distance}

\begin{figure}[ht]
\centering
\begin{minipage}[b]{0.45\linewidth}
\includegraphics[width=3.5in]{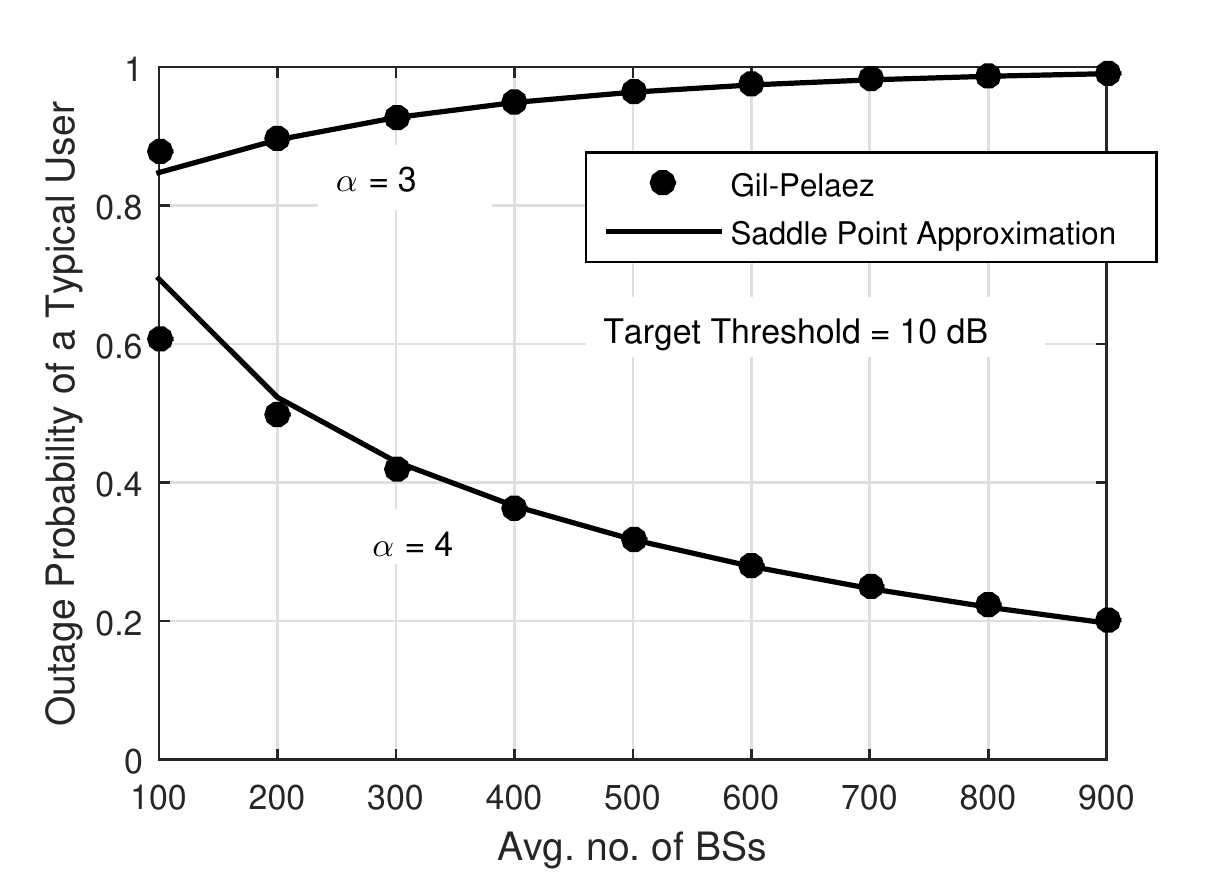}
	\caption{Case B: SIR outage vs. number of BSs at $\theta = 10$ dB, $P=0$ dB. }
	\label{fig:caseC-intensityTarget10dB}
\end{minipage}
\hfill
\begin{minipage}[b]{0.45\linewidth}
\includegraphics[width=3.5in]{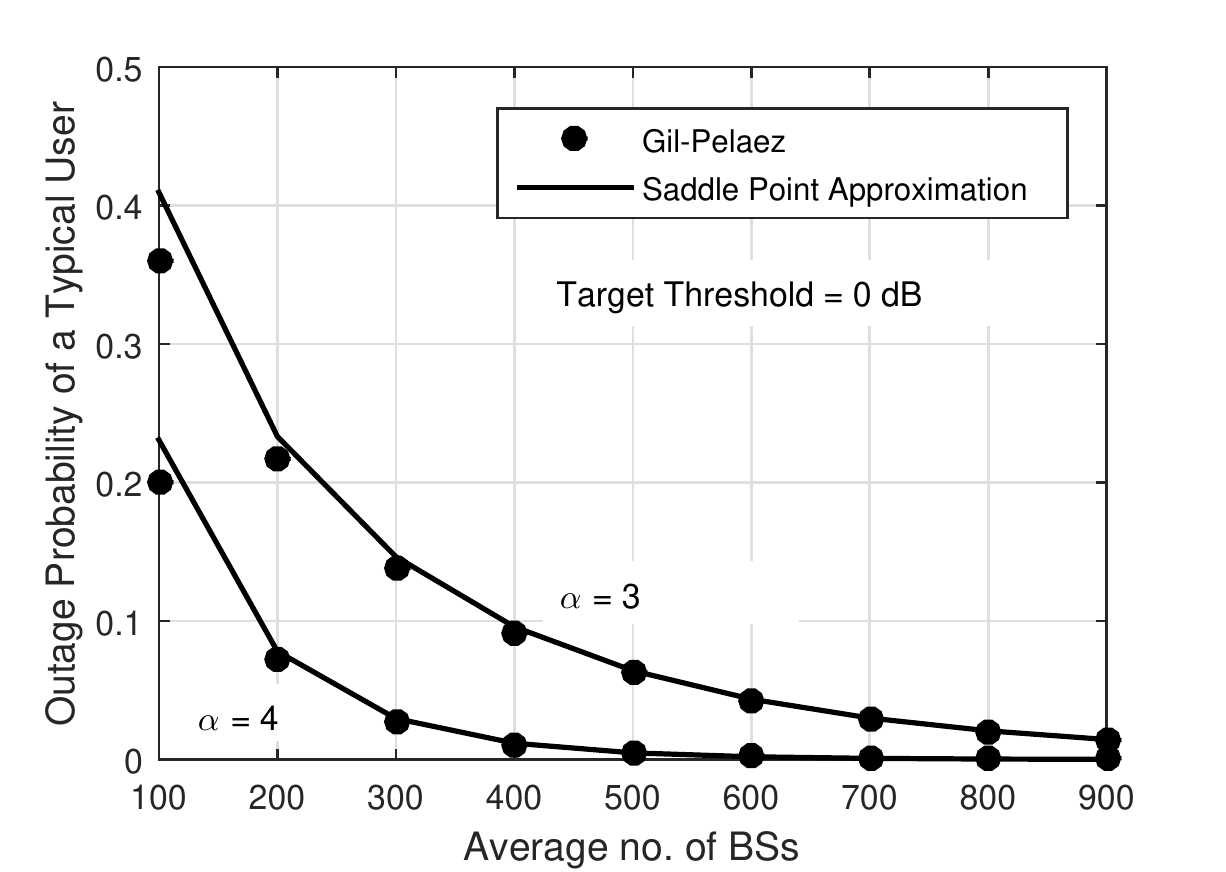}
	\caption{Case B: SIR outage vs. number of BSs at $\theta = 0$ dB, $P=0$ dB. }
	\label{fig:caseC-intensityTarget0dB}
\end{minipage}
\end{figure}

In both the Figs. \ref{fig:caseC-intensityTarget10dB} and \ref{fig:caseC-intensityTarget0dB}, we plot the outage probability versus the average number of the BSs in the system. For these plots, the radius of exclusion region is $a = 30$~m while the radius of cooperation was $R = 150$~m. The BSs are scattered in a uniform random manner over a total area of radius $1000$~m, such that the average number of BSs in the total area is fixed. The transmit power of the BSs is taken to be $P = 0$ dB. The target threshold SIR was maintained at $10$ dB for Fig. \ref{fig:caseC-intensityTarget10dB} and $0$ dB for Fig. \ref{fig:caseC-intensityTarget0dB}. The BS intensity is made high to ensure Gaussianity, as given by Corollary 2. Thus, we see that the outage probability given by Gil-Pelaez formula matches with those given by SPA as the intensity of BSs increases in both figures. We see that assuming different path-loss exponents gives different trends, depending on the target threshold. For $\alpha=4$, the outage decreases as the BS intensity increases for both Figs. \ref{fig:caseC-intensityTarget10dB} and \ref{fig:caseC-intensityTarget0dB}. Thus, the trend is independent of the target threshold. For $\alpha=3$, the outage increases as the BS intensity increases in Fig. \ref{fig:caseC-intensityTarget10dB}, while the outage decreases at lower threshold for Fig. \ref{fig:caseC-intensityTarget0dB}. This means that at a lower path-loss exponent, the interference does not attenuate fast enough so that the signals may dominate with increasing number of BSs, when the threshold is made high. Thus, the performance degrades with increasing BS intensity. The opposite is true for higher path-loss. This has an important practical implication, in that, the cooperative communication is viable in ultra dense networks only when the path-loss exponent is sufficiently high. 

\begin{figure}[h]
\begin{center}
	\includegraphics[width=4in]{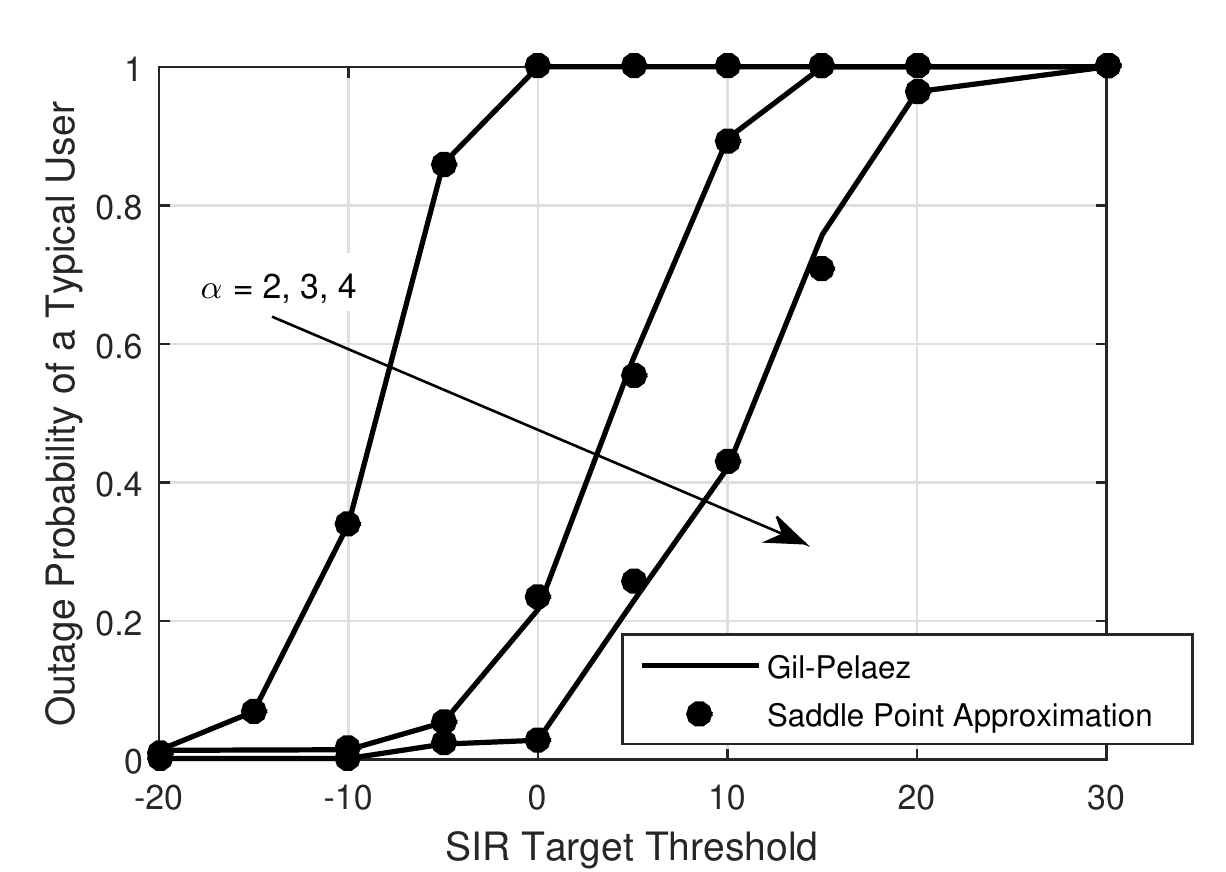}
	\caption{Case B: SIR outage vs. target threshold, $P$ = 0 dB, avg. no. of BSs $=200$. }
	\label{fig:caseC-target}
 \end{center}
\end{figure}

Finally, Fig. \ref{fig:caseC-target} plots the outage probability against the SIR target threshold for path-loss exponents $\alpha = 2, 3, 4$. The average number of BSs is $200$. All the other parameters are maintained as before. We see that the values given by Gil-Pelaez  matches well with the values given by SPA. As expected, the outage increases with increasing threshold.  We also notice that, for fixed threshold, when $\alpha$ increases, the SIR outage probability decreases. Thus, we see that attenuation has the effect of diluting the interference and enhancing the overall performance of the CoMP system.


\section{Conclusion}
We have presented a method to calculate the SINR outage probability of a typical user in a large-scale cellular network provided the exact moment/cumulants of the signal and interference are known using orthogonal polynomials. Exact formulas have been derived for the cases when the base distributions follow Gaussian and Student's $t$-distribution and their associated orthogonal polynomials are Hermite polynomials and Krishnamoorthy polynomials. Next, we have studied how the SINR outage can be calculated  using the saddle point method when the CGF also exists. We have studied the general method of saddle point approximation where we can adjust the skewness and kurtosis of the base distribution. Numerical results have been presented to check the accuracy of the proposed SIR outage approximation methods. The presented techniques can potentially solve a number of similar outage problems. 

Some of the possible future research direction can be as follows:~(i)~finding appropriate transformation to convert the Hamburger problem into Hausdorff problem so as to take the advantage of Legendre, Chebyshev, and Gegenbauer polynomials, which are all special cases of Jacobi polynomials, (ii) analyzing the CoMP transmission for different diversity combining techniques such as selection combining or equal gain combining, (iii) instead of fixed radius of cooperation, for which the number of cooperating BSs are random, the case for fixed number of cooperating BSs can be investigated, and (iv)~to study the qualitative relationship between the saddle point and SINR outage probability.

 
 \appendices
 
 \section*{Appendix A}
\renewcommand{\theequation}{A.\arabic{equation}}
\setcounter{equation}{0}
Using the Campbell's formula (\ref{eqn:campbell-formula}) for signal $X$, evaluating the integral over the limits $a$ and $R$, we have 
\[ \kappa_n(X) = 2 \pi \lambda \mu_n(G) P^n \int_a^R r^{-n\alpha + 1} \ud r = \frac{2 \pi \lambda \mu_n(G) P^n}{n\alpha-2}(a^{-n\alpha+2} - R^{-n\alpha+2}) . \]
Similarly, for interference $Y$, the limits of integral are $R$ to $\infty$, so
\[ \kappa_n(Y) = 2 \pi \lambda \mu_n(G) P^n \int_R^\infty r^{-n\alpha + 1} \ud r = \frac{2 \pi \lambda \mu_n(G) P^n}{n\alpha-2} R^{-n\alpha+2}. \]
Substituting the expressions for $\kappa_n(X)$ and $\kappa_n(Y)$ in (\ref{eqn:nth-cumulant-gamma-generic-model}) of Proposition \ref{prop:cumulant-generic-model}, we obtain 
\[ \kappa_n(\Omega) = \frac{2 \pi \lambda \mu_n(G) P^n}{n\alpha-2} a^{-n\alpha+2} \left[ \theta^n \Big( \frac{R}{a} \Big)^{-n\alpha + 2} + ( -1)^n \Big\{ 1 - \Big( \frac{R}{a} \Big)^{-n\alpha + 2} \Big\} \right].  \]
Simplifying, we obtain the desired result.

\section*{Appendix B}
\renewcommand{\theequation}{B.\arabic{equation}}
\setcounter{equation}{0}
Let the $k$-th Hermite polynomial be given by $He_k(\omega) = \sum_{i=0}^k h_{ki} \omega^i$, then the value of $a_k$  in (\ref{eqn:orthogonal-moment-1}), becomes $a_k = \frac{1}{k!} \sum_{i=0}^k h_{ki}\mu_i(\Omega)$. Therefore, the PDF of $\Omega$ reconstructed using its moments, via standard normal base distribution, is
\[ f_\Omega(\omega) = \frac{1}{\sqrt{2\pi}} e^{-x^2/2} \left[ \sum_{k=0}^\infty \frac{1}{k!} \left( \sum_{i=0}^k  h_{ki} \mu_i \right) He_k(\omega) \right]. \]

The CDF can be obtained by integrating as $F_\Omega(\omega) = \int_{-\infty}^\omega f_\Omega(x) \ud x$. We first note that $\int_{-\infty}^0 x^n e^{-x^2/2} \ud x = (-1)^n 2^{\frac{n-1}{2}} \Gamma(\frac{n+1}{2})$, which can be evaluated by changing the variable to $y = \frac{x^2}{2}$. Thus, 
 \[ F_\Omega(0) = \frac{1}{\sqrt{2 \pi}} \sum_{k=0}^\infty a_k \int_{-\infty}^0 He_k(x) e^{-x^2/2} \ud x. \]
Putting $He_k(x) = \sum_{i=0}^k h_{ki} x^i$, we get
\[  F_\Omega(0) = \frac{1}{\sqrt{2 \pi}} \sum_{k=0}^\infty a_k \sum_{i=0}^k h_{ki} \int_{-\infty}^0 x^i e^{-x^2/2} \ud x  \nonumber =  \frac{1}{\sqrt{2 \pi}} \sum_{k=0}^\infty a_k \sum_{i=0}^k (-1)^i 2^{\frac{i-1}{2}} \Gamma\left(\frac{i+1}{2}\right) h_{ki}. \]
Thus, the outage probability is given by $Q_\Omega(0) = 1 - F_\Omega(0)$, where $F_\Omega(0)$ is given using $\mu_n(\Omega)$.

\section*{Appendix C}
\renewcommand{\theequation}{C.\arabic{equation}}
\setcounter{equation}{0}
Assuming the Student's $t$-distribution, the reconstructed PDF of $\Omega$ is
\[ f_\Omega(\omega) = \left(1+\frac{\omega^2}{v}\right)^{-(v+1)/2} \sum_{i=0}^{\lfloor \tfrac{v}{2} \rfloor} a_i T_i(\omega), \]
 where $a_k = \sum_{i=0}^k t_{ki}\mu_i(\Omega)$, as given by (\ref{eqn:orthogonal-moment-1}). 
Integrating the PDF to find the CDF, we have $F_\Omega(\omega) = \int_{-\infty}^\omega f_\Omega(x) \ud x$. We first note that 
\[\int_{-\infty}^0 x^n \left(1+\frac{x^2}{v} \right)^{-(v+1)/2} \ud x = (-1)^n \frac{v^{(n+1)/2}}{2} B\left(\frac{v - n}{2},\frac{1 + n}{2}\right) \qquad \mathrm{for} \qquad n< v, \] 
 where $B(\cdot, \cdot)$ is the Beta function. Therefore, 
 \[ F_\Omega(0) = \sum_{k=0}^{\lfloor v/2 \rfloor} a_k \int_{-\infty}^0 \left(1+\frac{x^2}{v} \right)^{-(v+1)/2} T_k(x) \ud x.  \]

Putting $T_k(\omega) = \sum_{i=0}^k t_{ki} \omega^i$ and integrating, we have
\[F_\Omega(0) = \sum_{k=0}^{\lfloor v/2 \rfloor} a_k \sum_{i=0}^k (-1)^i \frac{v^{(i+1)/2}}{2} B\left(\frac{v - i}{2},\frac{1 + i}{2}\right) t_{ki}. \]
 Thus, the outage probability is given by $Q_\Omega(0) = 1 - F_\Omega(0)$, where $F_\Omega(0)$ is calculated as above using $\mu_n(\Omega)$. When the parameter $v$ is assigned using moment match, we have for the Student's $t$-distribution $\mathrm{Ex. Kurt}(X) = 6/(v-4)$. Solving for $v$, we obtain the desired parameterization.

\section*{Appendix D}
\renewcommand{\theequation}{D.\arabic{equation}}
\setcounter{equation}{0}

The integral of the CGF is evaluated as \cite{Lowen1990}
\[ \int_a^b (e^{-t P r^{-\alpha}} - 1) r \ud r  = \frac{(tP)^{2/\alpha}}{\alpha}\left[ \Gamma\left(-\frac{2}{\alpha},tPb^{-\alpha}\right) - \Gamma\left(-\frac{2}{\alpha},tPa^{-\alpha}\right) \right] - \frac{b^2 - a^2}{2}, \]
where $\Gamma(a,z) = \int_z^\infty x^{a-1} e^{-x} \ud x$ is the upper incomplete Gamma function.\footnote{
Using the recurrence relation $\Gamma(a+1,z) = a\Gamma(a,z) + z^a e^{-z}$, the integral is often expressed as \cite[Eqn 19]{Lowen1990}
\[\int_a^b (e^{-t P r^{-\alpha}} - 1) r \ud r  = -\frac{1}{2}\left[ b^2(1-e^{-tPb^{-\alpha}}) - a^2(1-e^{-tPa^{-\alpha}}) + (tP)^{2/\alpha}\left[ \Gamma\left(1-\frac{2}{\alpha},tPb^{-\alpha}\right) - \Gamma\left(1-\frac{2}{\alpha},tPa^{-\alpha}\right) \right] \right].  \]
} 
In order to differentiate this integral $n$-times with respect to $t$, consider the $n$-th derivative of the first term:
\[ \frac{\ud^n}{\ud t^n} \frac{(tP)^{2/\alpha}}{\alpha} \Gamma\left(-\frac{2}{\alpha},tPb^{-\alpha}\right) = \frac{b^2}{\alpha}  \frac{\ud^n}{\ud t^n} (tPb^{-\alpha})^{2/\alpha} \Gamma\left(-\frac{2}{\alpha},tPb^{-\alpha}\right). \]
Put $g(t) = tPb^{-\alpha} = z$ and $f(z) = z^{2/\alpha} \Gamma(-2/\alpha,z)$. We have the $n$-th derivative of $f(z)$ with respect to $z$ as \cite[Eqn 8.8.16]{Olver2010} $f^{(n)}(z) = (-1)^n z^{2/\alpha - n} \Gamma\left(n-\frac{2}{\alpha}, z\right)$. Similarly, $g'(t) = Pb^{-\alpha}$, while $g^{(n)} = 0$ for all $n \geq 2$. Now, using Faa di Bruno's formula, $\frac{\ud^n}{\ud t^n} f(g(t)) = \sum_{k=1}^n f^{(k)}(z) B_{n,k}(g'(t),0,\ldots,0 ),$
where $B_{n,k}$ is the partial exponential Bell polynomial. Here $B_{n,k}(g'(t),0,\ldots,0 )$ is 0 if $k<n$ and $g'(t)^n$ if $k=n$. Hence, $\frac{\ud^n}{\ud t^n} f(g(t)) = f^{(n)}(z) \cdot (g'(t))^n = (-1)^n b^{-2} P^{2/\alpha} t^{2/\alpha - n} \Gamma\left(n-\frac{2}{\alpha},tPb^{-\alpha}\right)$. Therefore, we have 
\[ \frac{\ud^n}{\ud t^n} \frac{(tP)^{2/\alpha}}{\alpha} \Gamma\left(-\frac{2}{\alpha},tPb^{-\alpha}\right) = \frac{(-1)^n}{\alpha}  P^{2/\alpha} t^{2/\alpha - n} \Gamma\left(n-\frac{2}{\alpha},tPb^{-\alpha}\right).\]

 We will have similar result for $\frac{(tP)^{2/\alpha}}{\alpha} \Gamma\left(-\frac{2}{\alpha},tPa^{-\alpha}\right)$. Lastly, the derivatives of the constant last term will be zero. Putting everything together, we have our desired result.

 
 \section*{Acknowledgments}
The authors would like to thank Prof. Martin Haenggi for helpful discussions. This work was funded by the Natural Sciences and Engineering Research Council of Canada (NSERC).  


 \bibliographystyle{IEEE} 

\end{document}